\documentclass[aps,prd,
preprint,tightenlines,preprintnumbers,%
superscriptaddress,notitlepage,nofootinbib,eqsecnum,floatfix]{revtex4-1}

\usepackage{bm}
\usepackage{graphicx}
\usepackage[pdftex]{color}
\usepackage{amsmath, amssymb}
\usepackage{hyperref}
\usepackage{array}
\usepackage{subfigure}
\usepackage{booktabs}
\hypersetup{
    colorlinks=true,     
    linkcolor=blue,      
    citecolor=blue,      
    filecolor=blue,      
    urlcolor=blue        
}
\usepackage{rotating}
\usepackage{mathtools}
\usepackage{multirow}

\usepackage{dcolumn} 

\newcommand{\Tr}{\ensuremath{\mathop{\text{Tr}}}}     

\newcommand{\spose}[1]{\hbox to 0pt{#1\hss}}

\newcommand{\inapprox}{\mathrel{\spose{\lower 3pt\hbox{$\mathchar"218$}}
 \raise 2.0pt\hbox{$\mathchar"232$}}}

\usepackage{mathtools}
\DeclarePairedDelimiter{\evdel}{\langle}{\rangle}
\newcommand{\ev}{\evdel}

\usepackage{xspace} 

\begin{document}


\title{Finite-temperature phase structure of SU$(4)$ gauge theory with multiple fermion representations}



\author{Venkitesh Ayyar}
\affiliation{Department of Physics, University of Colorado, Boulder, Colorado 80309, USA}

\author{Thomas DeGrand}
\affiliation{Department of Physics, University of Colorado, Boulder, Colorado 80309, USA}

\author{Daniel C. Hackett}
\affiliation{Department of Physics, University of Colorado, Boulder, Colorado 80309, USA}

\author{William~I.~Jay}
\affiliation{Department of Physics, University of Colorado, Boulder, Colorado 80309, USA}

\author{Ethan T.~Neil}\email{ethan.neil@colorado.edu}
\affiliation{Department of Physics, University of Colorado, Boulder, Colorado 80309, USA}
\affiliation{RIKEN-BNL Research Center, Brookhaven National Laboratory, \\ Upton, New York 11973, USA}

\author{Yigal~Shamir}
\affiliation{Raymond and Beverly Sackler School of Physics and Astronomy,
Tel~Aviv University, 69978 Tel~Aviv, Israel}

\author{Benjamin Svetitsky}
\affiliation{Raymond and Beverly Sackler School of Physics and Astronomy,
Tel~Aviv University, 69978 Tel~Aviv, Israel}

\date{\today} 

\begin{abstract}
We investigate the phase structure of SU(4) gauge theory with the gauge field simultaneously coupled to two flavors of fermion in the fundamental representation and two flavors of fermion in the two-index antisymmetric representation.
We find that the theory has only two phases, a low-temperature phase with both species of fermion confined and chirally broken, and a high-temperature phase with both species of fermion deconfined and chirally restored.
The single phase transition in the theory appears to be first order, in agreement with theoretical predictions.
\end{abstract}


\maketitle

\begin{flushleft}

\end{flushleft}

\tableofcontents

\section{Introduction}
	\label{sec:intro}
	We have been performing numerical simulations of SU(4) gauge theory coupled to
two flavors of  Dirac fermion in the fundamental (quartet) representation and two flavors
of Dirac fermion in the two-index antisymmetric (sextet) representation, which is a real representation.
These studies have been motivated by the use of a related theory---with three flavors of 
Dirac fundamentals and five Majorana sextets---as a model for a composite Higgs boson alongside a partially composite top quark \cite{Ferretti:2013kya,Ferretti:2014qta}. 
Our previous work \cite{Ayyar:2017qdf,Ayyar:2018zuk} was concerned with the mesonic and baryonic properties of this system.
Here we describe our studies of its finite temperature behavior.

The presence of multiple fermion representations (a {\em multirep\/} theory) opens the possibility of dynamical scale separation between the confinement and chiral transitions for each representation.
Scale separation may arise from a variety of different mechanisms.
One possibility is separation between the chiral symmetry breaking scales associated with the two representations, as in the ``tumbling" scenario \cite{Raby:1979my,Marciano:1980zf} in which chiral symmetry is broken spontaneously at a hierarchy of different mass scales. 
Quenched studies on very small lattices in the early 80's indicated the existence of 
separated chiral transitions for different fermion representations \cite{Kogut:1983sm,Kogut:1982rt,Kogut:1984nq,Kogut:1984sb},
but it is not known whether these results persist in the presence of dynamical fermions.
It also may be possible for the scales of the chiral and confinement transitions to be 
different.
Previous work with dynamical fermions has pointed to this 
possibility \cite{Karsch:1998qj}, but it is possible that the theory explored in that work---the SU(3) theory with adjoint fermions---is infrared conformal rather than confining \cite{*[{The SU(2) theory with adjoint fermions has been studied at length and found to be conformal.  See for example }] [{ and references therein.}] Rantaharju:2015cne}.
A final possibility is that the confinement transitions of different representations can be separated: if center symmetry breaks in several stages, there may exist phases where some representations of charge are deconfined while others remained confined. 

Our numerical data lead us to the conclusion that 
there is only a single  finite-temperature phase transition in this theory, 
with the characteristics of chiral restoration and deconfinement for both fermion species.
We find this to hold in 
limit theories where one or the other fermion species is decoupled, as well as in the full theory coupled to both species simultaneously.
The sextet-only theory has an order parameter which characterizes its phase,
and so it presumably possesses a real phase transition line. We could not determine its order.
The fundamental-only theory has no order parameter for confinement and appears to exhibit 
crossover behavior.
For the full theory, we find only first-order transitions in the region we explore.
We summarize our knowledge in the ``Columbia plots'' (Figs.~\ref{fig:theory-columbia-plot} and \ref{fig:col_plot}) below.

The paper is organized as follows:
In Section \ref{sec:theory}, we discuss the theory, its symmetries, and the expected 
behavior of its confinement and chiral transitions.
We discuss our lattice methodology in Section~\ref{sec:lattice_methods}.
We then present our numerical findings.  In Section \ref{results_limiting_cases}, we 
examine the phase structure in two limiting cases of the theory, keeping
only fundamental fermions, or only sextet ones.
We then examine the phase structure
of the full theory with both species of fermion in Section~\ref{sec:phase}.
We conclude in Section~\ref{sec:conclusions}. A preliminary account of this work can be found in \cite{Ayyar:2017uqh}. For lattice work on a composite Higgs model with a different gauge group, see \cite{Bennett:2017kga}.

In what follows, a quantity labeled $m_4$, $P_4$, etc., corresponds to that quantity as measured for the fundamental fermions, while $m_6$, $P_6$, etc., correspond to the sextet fermions.

\section{Continuum Theory}
	\label{sec:theory}
\subsection{Polyakov Loops and Multiple Representations}
\label{subsec:hrpl}

The usual diagnostic for confinement is the Polyakov loop, which can be constructed from gauge links in any representation of the group.
For our study, the relevant Polyakov loops are defined as
\begin{align}
	P_4(\vec{x}) &= \Tr \Omega_{\vec{x}} \\
	P_6(\vec{x}) &= \tfrac{1}{2} \left[
		(\Tr \Omega_{\vec{x}})^2 - \Tr (\Omega_{\vec{x}})^2
	\right]
\end{align}
where
\begin{equation}
	\Omega_{\vec{x}} = \prod_{t=1}^{N_t} U_{\hat{t}}(\vec{x}, t).
\end{equation}
In SU(3), it is possible to write any higher-representation Polyakov loop in terms of the fundamental Polyakov loop and its complex conjugate.
The behavior of higher-representation Polyakov loops is thus completely determined if the fundamental Polyakov loop is known.
In SU(4), $P_4$ and $P_6$ are a sufficient set.

The expectation value of a Polyakov loop in a representation $R$ measures the free-energy $F_R$ of a static charge in that representation via
\begin{equation}
|\ev{P_R}| \sim \exp[-F_R/T].
\end{equation}
This Polyakov loop is an order parameter if the fermion action preserves enough center symmetry (invariance under global $Z_N$ gauge transformations) to protect it in the unbroken phase.
In an SU(4) gauge theory, sextet fermions break $Z_4$ to $Z_2$.
The residual $Z_2$ symmetry is not enough to protect the sextet Polyakov loop, but the fundamental Polyakov loop remains an order parameter when only sextet fermions are present \cite{DeGrand:2015lna}.
Adding fundamental fermions breaks the $Z_4$ center symmetry completely, so in the full theory neither Polyakov loop is an order parameter.

Even when Polyakov loops are not order parameters, they are observed to jump from a small value to a large value as the temperature rises.
This jump signals a qualitative change in the physics of color charge screening.
It could be that this occurs at different temperatures for fermions in different representations.

\subsection{Order of Chiral Phase Transitions}
\label{sec:pisarski_wilczek}

Each representation of fermion could have its own distinct chiral transition.
There are three possible orders in which the chiral transitions may be encountered
when cooling from $T=\infty$ to $T=0$: sextet first (as predicted by
arguments along the lines of the ones in Ref.~\cite{Raby:1979my}), 
or fundamental first, or a simultaneous transition.
As discussed below in Section~\ref{sec:phase}, our results indicate that
the third possibility is what occurs for our system.

One of us \cite{Hackett:2017gti} has carried out a Pisarski--Wilczek stability analysis \cite{Pisarski:1983ms} for this single phase transition, based on the pattern of chiral symmetry breaking of the two fermion species.
This method consists of examining the critical behavior of a three-dimensional effective theory of the chiral condensates of the theory---a linear sigma model.
If the effective theory has any stable fixed points, it may undergo a second-order phase transition governed by one of the fixed points, but the transition may also be first-order.
In the absence of stable fixed points, the analysis predicts a first-order transition.
This procedure generalizes straightforwardly to theories where there are multiple 
representations of fermion, as long as the chiral symmetry breaking pattern can be written as a direct product of the patterns of the single-representation sectors (up to additional U(1) factors due to non-anomalous axial symmetries).
The effective theory for the multirep theory is then simply the single-representation theories coupled together.
The prediction of this calculation \cite{Hackett:2017gti}, carried out to first order in $\epsilon=4-d$, is a first-order phase transition.

Fig.~\ref{fig:theory-columbia-plot} is a rough sketch of a ``Columbia plot'' summarizing the theoretical predictions for the nature of the finite-temperature transition in the various fermion-mass regimes (with some inputs from our results discussed below).
This sketch is made in analogy with the QCD Columbia plot, where the order of the phase transition encountered is plotted as a function of $m_u=m_d$ and $m_s$. Here, we plot the order of the phase In the pure-gauge limit, the transition is first order
{\cite{*[{See for example }] [{ and references therein.}] Datta:2009jn}.
The stability analysis predicts that the transition in the double chiral limit $m_4=m_6=0$ will also be first order.
First-order transitions are generically robust against small perturbations, so these transitions presumably extend into the regions around $m_4=m_6=\infty$ and $m_4=m_6=0$.
High-order Pisarski--Wilczek calculations \cite{Basile:2004wa,Basile:2005hw} indicate that the transitions in the massless limits of the fundamental-only and sextet-only theories can be second-order
[$(m_4,m_6)=(0,\infty)$ or~$(\infty,0)$].
The fundamental-only limit, SU(4) with two flavors of fundamental fermion, 
is similar to QCD with $m_u=m_d=0$ and $m_s=\infty$.  QCD in this limit is
believed to exhibit a second order phase transition with $O(4)$ critical exponents
(compare the discussion in Ref.~\cite{Soltz:2015ula}),
 so we expect the fundamental-only theory to behave similarly.
Fig.~\ref{fig:theory-columbia-plot} shows a second-order phase transition in either of these limits.
In the single-species limits, nonzero fermion mass will convert a second-order  chiral
transition into a  crossover.
Adding heavy fermions of the other species can leave the second-order transition undisturbed or convert it to first order, as shown, but the phase transition cannot disappear as long as one species is exactly massless.
If either single-species transition were first order,
 there would be a first order region in the corresponding corner.
In the pure sextet theory there will be a confinement transition for all values of $m_6$ because of the residual $Z_2$ center symmetry; we indicate a second-order transition there, though it could be first-order.

There are no analytical predictions for the intermediate region,
 where $m_4$ and $m_6$ are neither light nor heavy; 
we thus have no predictions for whether the first order regions connect, or whether there is an intermediate continuous crossover region.

\begin{figure}
	\centering
	\includegraphics[]{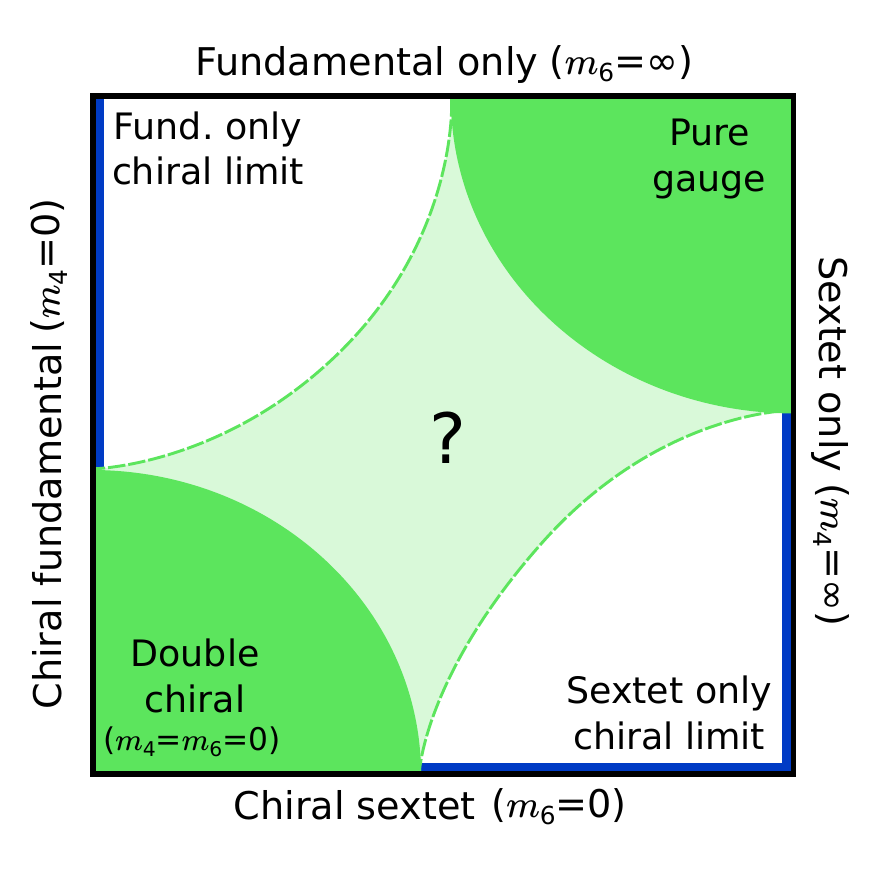}
	\caption{
		Columbia plot illustrating expectations for the order of the finite-temperature phase transition.
		The axes are the masses of the two fermion species in the theory, with $m_4$ on the x-axis and $m_6$ on the y-axis.
		The upper right corner is the pure-gauge limit; the lower left corner is the double chiral limit; the upper-left corner is the fundamental-only chiral limit; the lower-right corner is the sextet-only chiral limit.
		Green fields indicate regions of parameter space where the theory is predicted to exhibit a first order transition. Blue lines indicate regions of parameter space where the theory is predicted to exhibit a second-order transition.
	}
	\label{fig:theory-columbia-plot}
\end{figure}

\section{Lattice methods}
	\label{sec:lattice_methods}
	\subsection{Action}
\label{subsec:action}

For this study, we employ the same lattice action as in Ref.~\cite{Ayyar:2017qdf}.
For the fermions, we use a clover-improved Wilson action built from fat gauge links constructed by normalized-hypercubic (nHYP) smearing \cite{Hasenfratz:2001hp,Hasenfratz:2007rf}.
We construct the action for the sextet fermions by promoting the smeared links to the sextet representation \cite{DeGrand:2015lna}.
We set both clover coefficients equal to unity, $c_\text{SW} = 1$, a choice motivated by results from Ref.~\cite{Shamir:2010cq}.
The action for the gauge sector is the usual plaquette action with gauge coupling $\beta$ augmented by an nHYP-dislocation suppression (NDS) term \cite{DeGrand:2014rwa}, constructed from the nHYP-smeared links.
As in our zero-temperature study, we fix the NDS parameter $\gamma$ such that $\beta/\gamma=125$.
Altogether, the simulation parameter space is three dimensional: $\beta$, and two hopping parameters  $\kappa_4$ and $\kappa_6$.

\subsection{Spectroscopy}
\label{sec:spectroscopy}

Fermion masses $m_4$ and $m_6$ for the two representations
 are defined through the axial Ward identity (AWI),
\begin{equation}
\partial_\mu \langle 0 | A_{\mu a}^{(r)}(x) \mathcal{O}_r(0) | 0 \rangle
= 2 m_r \langle 0 | P_a^{(r)}(x) \mathcal{O}_r(0) | 0 \rangle,
\label{eq:pcac_continuum}
\end{equation}
where $a$ is an isospin index.
We use the local unimproved axial current $A_{\mu a}^{(r)}$ and pseudoscalar
 density $P_a^{(r)}$ in each representation $r$.
For $\mathcal{O}_r$ we take a pseudoscalar source.
The axial Ward identity is a statement of current conservation and is thus local and relatively insensitive to finite-volume effects as long as we stay in a confined phase.

We extract meson screening masses in the scalar, pseudoscalar, vector, and pseudovector channels from two-point correlation functions extending in a spatial lattice direction.
This is a standard technique in a finite temperature simulation.
We construct propagators with composite boundary conditions to double the effective 
length of the lattice \cite{Blum:2001xb,Aoki:2005ga,DeGrand:2007tx}.
These are called ``P+A correlators'' in the literature.

\subsection{Data sets}

In order to search for the various possible phase transitions, we explored a wide region of the three-dimensional bare parameter space.
This required an unusually large and heterogeneous data set, summarized in Table~\ref{tab:finite-t-dataset}.
To render this exploration tractable, we found it necessary to automate much of our data generation and analysis; see \cite{taxi-proc} for a description of our methods. 

For the full theory we focused predominantly on $\beta = 7.4$ and $\beta=7.75$,
 mostly on $ 12^3 \times 6 $ and $ 16^3 \times 8 $ volumes but with some additional data
 on $18^3 \times 6$ and $24^3 \times 8$ to check for finite-volume effects.
For the single representation theories, we ran only on a $12^3 \times 6$ volume.
We also made use of zero-temperature data to determine the lattice scale,
the fermion masses, and the pseudoscalar-to-vector mass ratio for some bare couplings near the transition.
Table~\ref{tab:zero-t-dataset} in the Appendix summarizes these zero-temperature data sets.
We use the lattice scale to derive the physical temperature along the phase boundary.

\begin{table}
	\centering
    \begin{ruledtabular}
	\begin{tabular}{lrlc}

		Theory & Volume & Subset & Ensembles \\
		\hline
		fundamental-only & $12^3\times6$ &              & 121 \\
		sextet-only & $12^3\times6$ &              & 239 \\
		full theory & $12^3\times6$ & $\beta=7.4$  & 128 \\
		& 				 & $\beta=7.75$ & 135 \\
		& 				 & 			All & 409 \\
		& $16^3\times8$ & $\beta=7.4$  &  22 \\
		&				 & $\beta=7.75$ &  35 \\
		& 				 & 			All &  57 \\
		& $18^3\times6$ & 			    &  49 \\
		& $24^3\times8$ & 			    &  26 \\
	\end{tabular}
    \end{ruledtabular}
	\caption{
		Summary of finite-temperature ensembles.
	}
	\label{tab:finite-t-dataset}
\end{table}

\subsection{Phase diagnostics}

With two species of fermion, there are (in principle) four distinct transitions to be considered, namely, the confinement and chiral transitions for each representation.
We need independent observables for each of these.

\subsubsection{Confinement transition}
\label{sec:conf_diag}

Polyakov loops in the  fundamental and sextet  representations are used
 to tell whether each fermion representation is confined or deconfined.
We also employ two additional quantities based on the Wilson flow \cite{Narayanan:2006rf,Luscher:2009eq}: 
the flowed anisotropy and Polyakov loops at long flow time.

The flowed observable $\ev{t^2 E(t)}$, where $ E $ is the energy density at flow time $t$, is commonly used to determine the scale for zero-temperature lattices.
$E$ is typically defined by summing over all orientations of clover terms or of plaquettes.
When measured on finite temperature lattices, spatial--temporal anisotropy 
in $\ev{t^2 E(t)}$ can be employed to determine the phase \cite{Datta:2015bzm,Datta:2016kea,Wandelt:2016oym}. 
Previous applications of anisotropy  have used the quantity
\begin{equation}
\Delta(t) = t^2 \langle E_{ss}(t) - E_{st}(t) \rangle,
\label{eqn:flow-split-obs}
\end{equation}
where $ E_{ss}(t) $ and $ E_{st}(t)$ represent the contribution from space--space and space--time clovers.
In this work we look at a related quantity, the flowed anisotropy $R_E$, defined as
\begin{equation}
	R_E(t) = \frac{\ev{E_{ss}(t)} }{ \ev{E_{st}(t)}} = 1+
\frac{ \Delta(t) }{ \langle t^2 E_{st}(t) \rangle }
\end{equation}
(cf.~Ref.~\cite{Borsanyi:2012zr}).
In the low-temperature confined and chirally broken phase, the gauge fields are roughly isotropic and the observable $R_E(t) \approx 1$ for any reasonable flow time $t$.
In the high-temperature deconfined and chirally restored phase, hypercubic symmetry is broken
 strongly: the temporal center symmetry is broken, while the spatial center 
symmetries are still (almost) preserved.
In such anisotropic phases, $R_E(t)$ departs from unity even at small flow times.
In this paper, we always measure $R_E(t)$ at flow time $t/a^2 = 1$.

The behavior of Polyakov loops at long flow times provides a sharp diagnostic of confinement.
For a lattice with temporal extent $ N_t $, the flow time
 ratio $ c_t = \sqrt{ 8 t } / (N_ta) $ is a rough measure of the
 Wilson-flow smearing in the temporal direction. 
Defining ``long flow time'' as $c_t > 1$, we find that flowed Polyakov loops 
exhibit nearly binary behavior at long flow times, depending on the phase.
On deconfined configurations, volume-averaged Polyakov loops rapidly reach their maximal values $\max |P_R| = d(R)$, where $d(R)$ is the dimension of the representation $R$.
On confined lattices, volume-averaged Polyakov loops wander or move only
 very slowly towards their maximal values.

All of our phase diagnostics (unflowed Polyakov loops, flowed
anisotropy, Polyakov loops at long flow time) agree everywhere in our data set, within our resolution in coupling space.
The flowed anisotropy and flowed Polyakov loops may be used to determine the phase of an ensemble without comparing it with nearby ensembles or picking some arbitrary threshold value, as is required when using unflowed Polyakov loops.
Such ensemble-local observables are better suited for automation.

For further discussion of these flow-based diagnostics, see Ref.~\cite{Ayyar:2017vsu}.

\subsubsection{Chiral transition}
\label{sec:Chiral-diagnostics}

Because we are using Wilson fermions, we 
use an indirect probe to determine
 whether chiral symmetry is broken in each sector: parity doubling in the meson sector.
In the chirally restored phase we expect parity partners to be degenerate.
We thus examine the mass splittings between the scalar and pseudoscalar states and between the 
vector and axial vector states, for each species of fermion.

\section{Phase Structure of Limiting Cases}
	\label{results_limiting_cases}

	\subsection{Sextet-Only Theory: $\kappa_4=0$}
    	\label{sec:sextet-only}
    	\subsubsection{Phase structure}

\begin{figure}
	\includegraphics{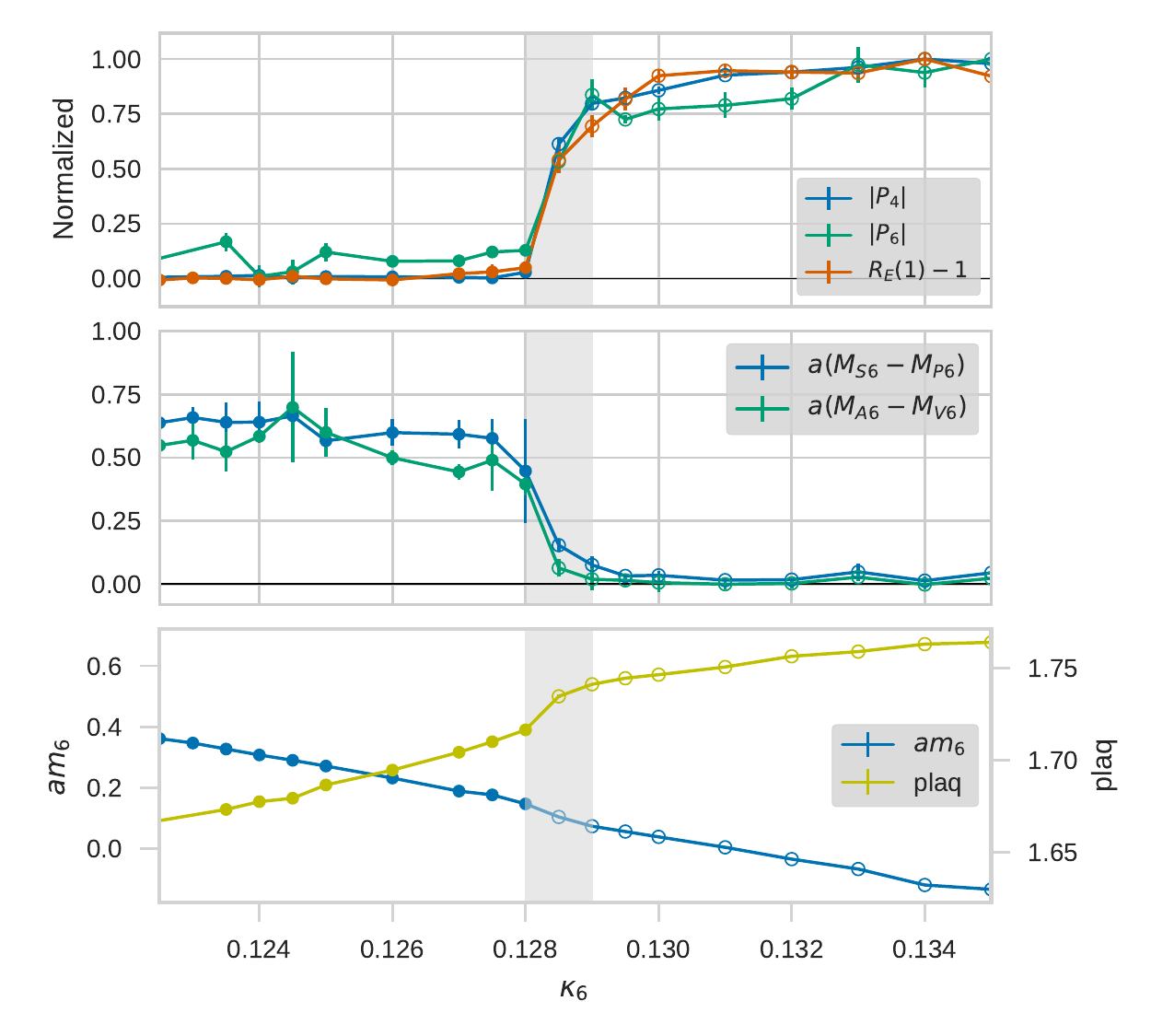}
	\caption{
		The variation with $\kappa_6$ of various quantities in the sextet-only theory for $\beta=8.5$ on $12^3 \times 6$.
		The top panel shows diagnostics of confinement: unflowed Polyakov loops and the flowed anisotropy at $t/a^2=1$.
		(Quantities are normalized by their maximum values along the slice for ease of comparison of their qualitative behavior.)
		The middle panel shows chiral diagnostics, the mass splittings of parity-partner mesons.
		The bottom panel shows the plaquette and the AWI mass.
		Points with closed (open) circles are deemed confined (deconfined) according to the behavior of Polyakov loops at long flow time.
		There is a single transition (gray band) from the confined and chirally broken phase to the deconfined and chirally restored phase.
		}
	\label{fig:sextet-slice}
\end{figure}

We begin with the gauge theory coupled only to sextet fermions.
Figure~\ref{fig:sextet-slice} shows the behavior of our various observables 
 along a typical slice through bare parameter space.
The top panel shows the behavior of the fundamental and sextet Polyakov loops.
Dynamical sextets break the center symmetry from $Z_4 \rightarrow Z_2$; the fundamental loop
 $P_4$ is an order parameter for the spontaneous breaking of the
 residual $Z_2$ center symmetry \cite{DeGrand:2015lna}
 and thus for confinement of static charges in the fundamental representation.
Although it is not an order parameter, we also examine the sextet Polyakov loop.
In the top panel, we see that all three  confinement diagnostics jump simultaneously and vary only smoothly elsewhere: there is only a single confinement transition in this theory.
The middle panel shows the  mass splittings of the sextet mesons.
The parity partners become degenerate simultaneously, indicating the restoration of chiral symmetry.
Within our resolution in $\kappa_6$, 
the confinement transition coincides with the chiral transition.

\begin{figure}[htb!]
	\includegraphics{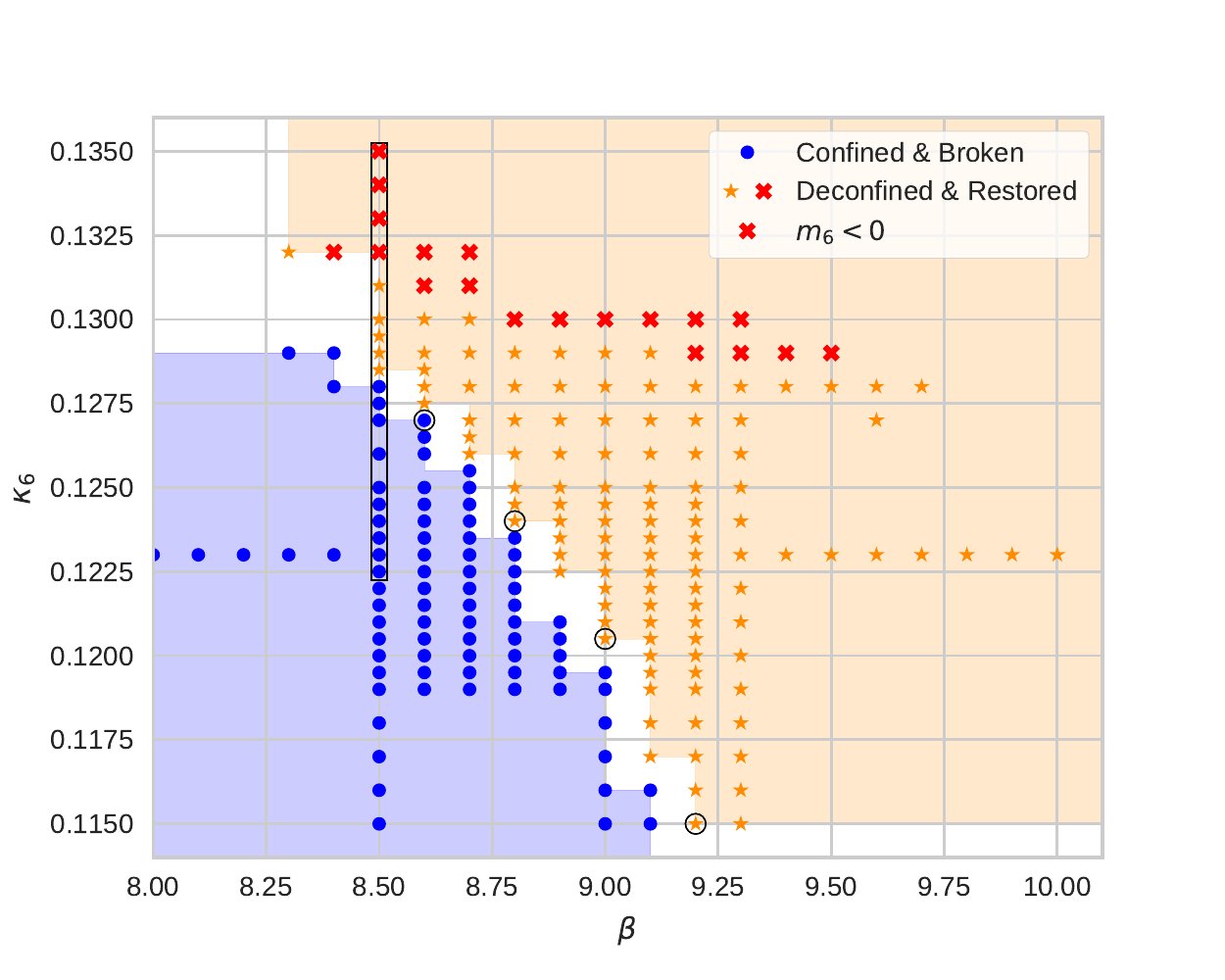}
	\caption{
		Phase diagram for the sextet-only limit  for $ 12^3 \times 6 $ lattices.
		Blue dots are confined and chirally broken ensembles while yellow stars and red Xs indicate deconfined and chirally symmetric ensembles.
		The blue field thus indicates the confined and chirally broken region of parameter space, while the orange field indicates the deconfined and chirally restored region of parameter space.
		The red Xs mark deconfined ensembles where $m_{6}<0$.
		The black box indicates the slice through bare parameter space shown in Fig.~\ref{fig:sextet-slice}.
		The circled ensembles have matching zero-temperature ensembles.
	}
	\label{fig:sextet-phase-diagram}
\end{figure}

The behavior shown on the slice in Fig.~\ref{fig:sextet-slice} is typical for this theory: everywhere we have looked in parameter space, we see a single, unified confinement and chiral transition, as in QCD\@.
Figure~\ref{fig:sextet-phase-diagram} summarizes our findings for the phase diagram for the sextet-only theory in the $\beta$--$\kappa_6$ plane.

For the four points marked by circles in Fig.~\ref{fig:sextet-phase-diagram}, we ran zero-temperature simulations at the same bare couplings in order to determine the lattice scale (see Table~\ref{tab:zero-t-dataset}).
As we describe in Appendix ~\ref{sec:scale-setting}, we set the lattice scale in each zero-temperature ensemble through calculation of the flow scale $t_1/a^2$.
Choosing the fiducial value $1/\sqrt{t_1}\equiv780$~MeV gives a physical value to the lattice spacing $a$ in each ensemble, and hence to the temperature $T=(N_ta)^{-1}$.
As can be seen in Fig.~\ref{fig:sextet-phase-diagram}, one of the ensembles is a blue point on the confined side of the transition, while the other three are orange points on the deconfined side.
These provide lower and upper bounds, respectively, on $T_c$ at the corresponding $m_6$ values.
We plot these temperatures in Fig.~\ref{fig:A2-Tc-vs-mq}.
The transition temperature curve must pass below the upper bounds (downward arrows) and above the lower bound (upward arrow).
We can compare the transition temperatures seen here to those in more familiar theories: the transition in the pure SU(3) gauge theory occurs near 280~MeV \cite{Lucini:2012wq}, while the crossover in QCD at physical quark masses occurs near 150~MeV \cite{Soltz:2015ula}.

\begin{figure}[htb!]
	\centering
	\includegraphics{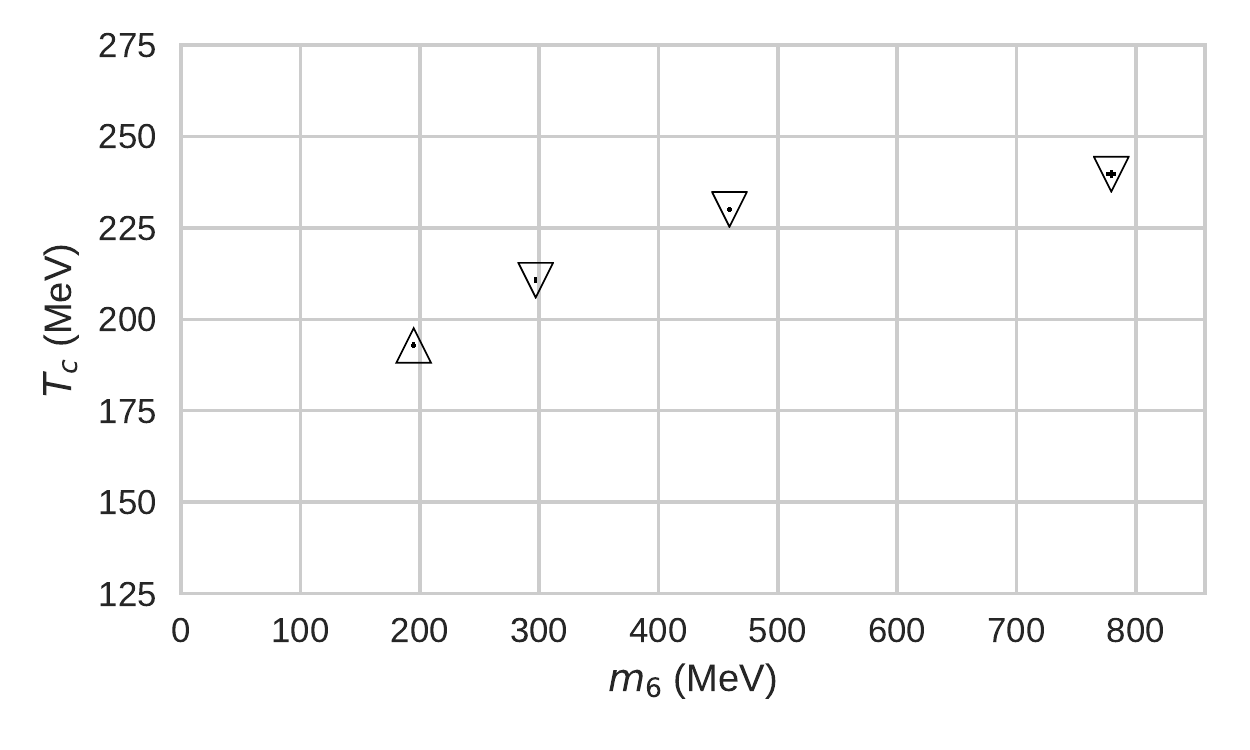}
	\caption{
		Transition temperature $T_c$ for the sextet-only theory with $N_t=6$ as a function of quark mass.
		The three heaviest ensembles (down arrows) are deconfined-side ensembles and thus give upper bounds on $T_c$.
		The lightest ensemble (up arrow) is a confined-side ensembles and thus gives a lower bound on $T_c$.
		The scale has been set via $t_1 \equiv 1/(780~\text{MeV})^2$.
	}
	\label{fig:A2-Tc-vs-mq}
\end{figure}

While the transition seen in Fig.~\ref{fig:sextet-slice} may look continuous, we can make no claim concerning the order of the transition for any value of $m_6$ since we have only a single volume.  As the fermion mass $m_6\to\infty$ we expect to obtain SU(4) pure gauge theory. In this limit there is a first order transition, which should persist for large values of $m_6$.

\subsubsection{Effects of the NDS action}

Our collaboration has previously examined the phase diagram of the sextet-only theory \cite{DeGrand:2015lna}, but without the NDS term in the gauge action.
In this previous study, the sextet-only theory was found to have a bulk transition.
The plaquette showed a large discontinuity at a $\kappa_6$ value below the thermal transition.
This large discontinuity is absent in our data. 
In Fig.~\ref{fig:sextet-slice}, the plaquette shows structure occurring simultaneously with the response of the Polyakov loops, but no additional structure.

To search for a bulk transition, we ran a grid of $4^4$ ensembles over the same region of bare 
parameter space covered by our $12^3\times6$ data. We found that the finite-temperature transition shifts substantially when changing $N_t$, thus confirming that it is not a softened bulk transition.
The NDS term appears to have completely banished the bulk transition, at least from the region of bare parameter space that we have explored.

	\subsection{Fundamental-Only Theory: $\kappa_6=0$}
		\label{sec:fund-only}
		
\begin{figure}
	\includegraphics{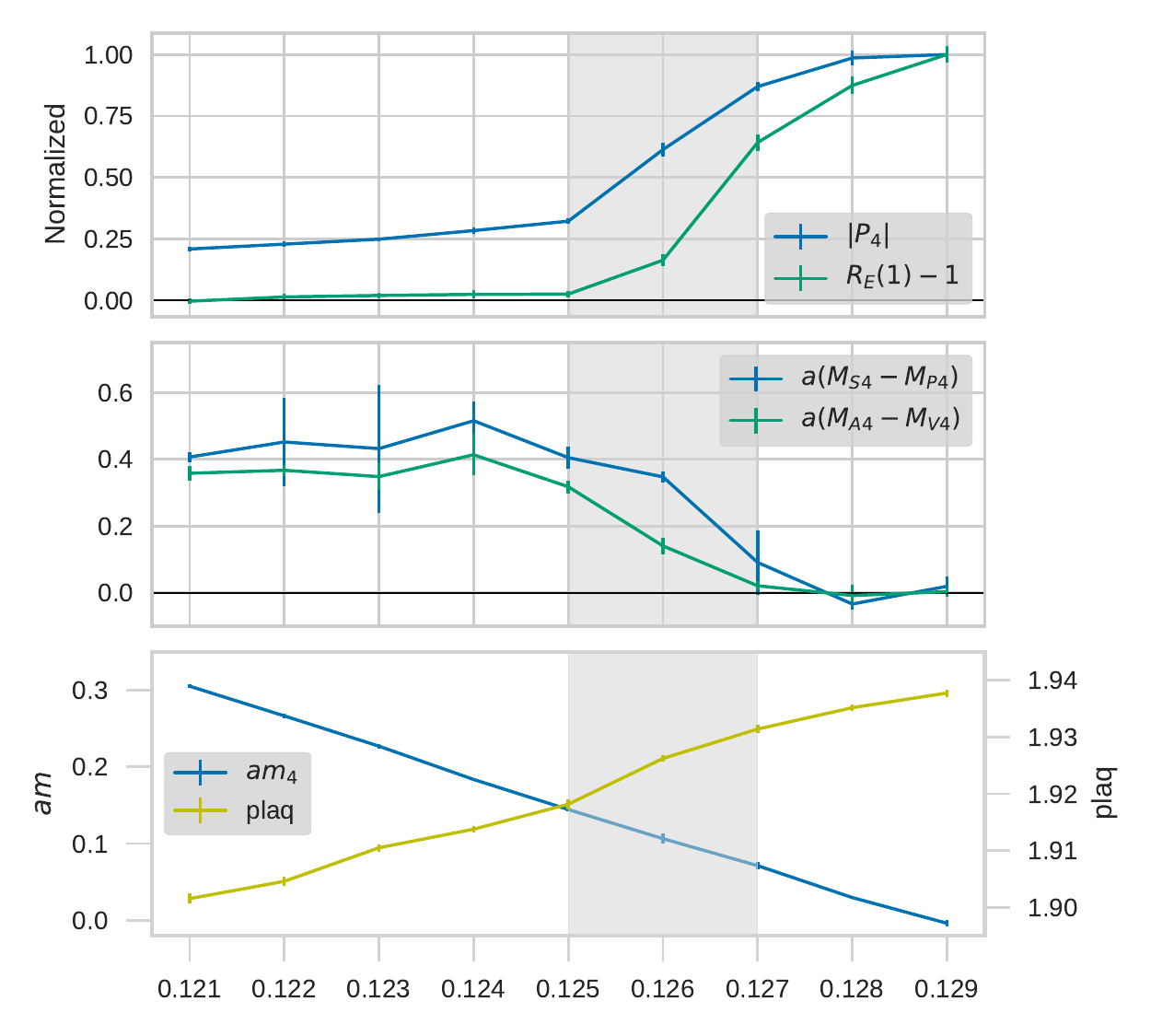}
	\caption{
		The variation with $ \kappa_4 $ of various quantities in the fundamental-only theory for $\beta=9.2$ on $12^3 \times 6$.
		The top panel shows diagnostics of confinement: the unflowed fundamental Polyakov loop and the flowed anisotropy at $t/a^2=1$.
		The middle panel shows diagnostics of the chiral transition---the mass splittings of parity-partner mesons.
		The bottom panel shows the plaquette and AWI mass.
		The peaks of the Polyakov loop and chiral susceptibilities lie somewhere in the gray band.
	}
	\label{fig:fund-slice}
\end{figure}

The other limiting case of our model contains only fundamental fermions and no sextets.
Figure~\ref{fig:fund-slice} shows the behavior of our  observables along a typical slice
 through bare parameter space, varying $\kappa_4$ over the transition at fixed $\beta=9.2$.
The top panel shows our confinement diagnostics, the fundamental Polyakov loop and the flowed anisotropy.
The two quantities change simultaneously and only once, varying smoothly otherwise: there is only a single crossover in each observable.
The middle panel shows our chiral diagnostics, the mass splittings of parity-partner mesons.
The splittings smoothly go to zero beyond $ \kappa_4 = 0.127 $, indicating chiral restoration. 
Comparing the top and middle panels, we see that the chiral and confinement crossovers overlap.
In the bottom panel, we see that the quark masses and plaquette vary smoothly, as expected for a crossover.

\begin{figure}[htb]
	\includegraphics{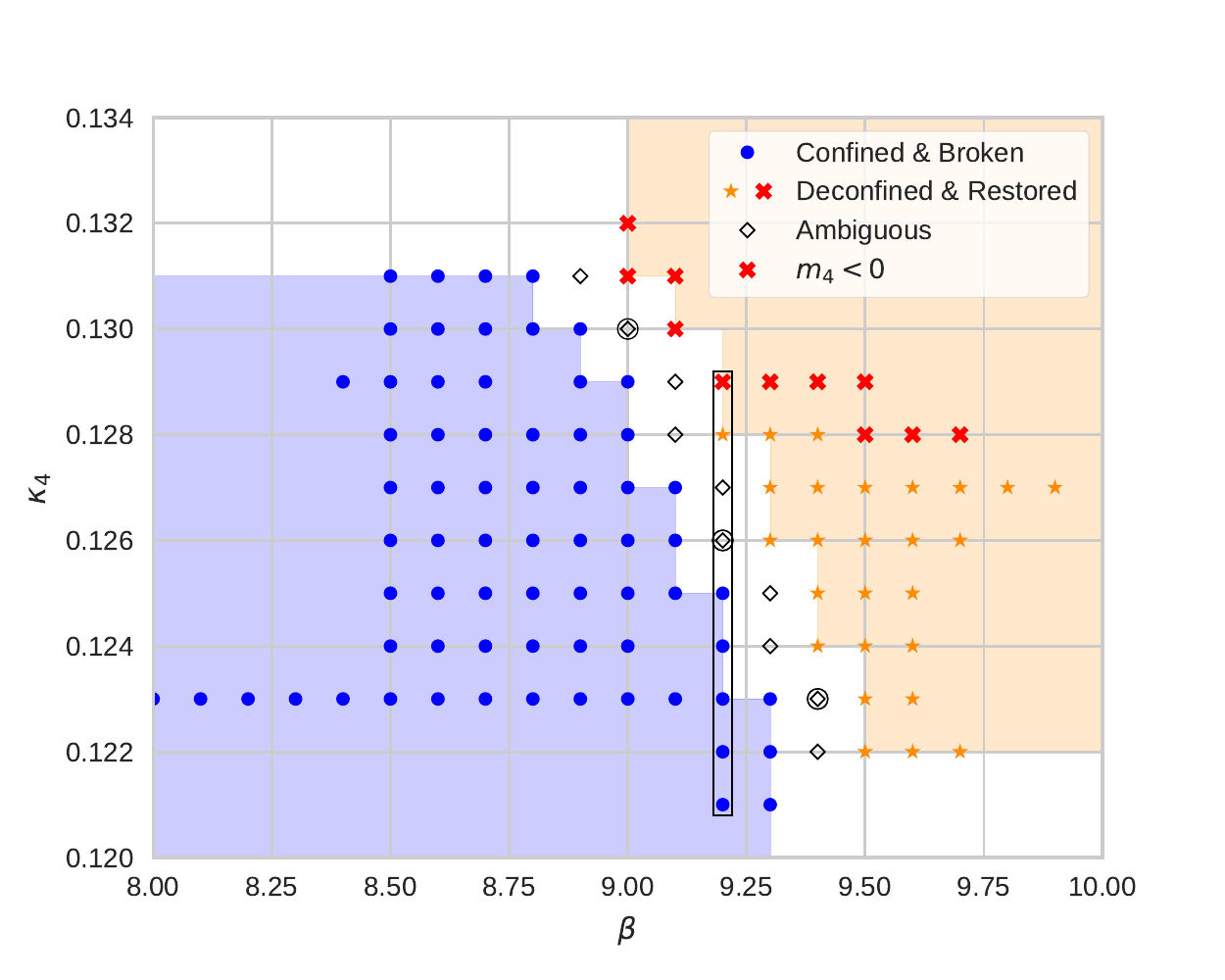}
	\caption{
		Phase diagram for the fundamental-only limit on $12^3 \times 6$.
		Symbols and colors are as in Fig.~\ref{fig:sextet-phase-diagram}, with the addition of hollow diamonds indicating ensembles in the crossover region, where
		the diagnostics are ambiguous.
		The black box indicates the slice through bare parameter space shown in Fig.~\ref{fig:fund-slice}.
		The circled ensembles have matching zero-temperature ensembles available.
	}
	\label{fig:fund-phase-diagram}
\end{figure}

The behavior observed on the slice in Fig.~\ref{fig:fund-slice} is typical for this theory: everywhere we have investigated, we observe only a single unified chiral and confinement crossover.
Figure~\ref{fig:fund-phase-diagram} summarizes our findings for the $\beta$--$\kappa_4$ phase diagram for the fundamental-only theory.
As we did for the sextet theory, we have determined the physical temperature at three points along the transition, this time choosing three points inside the crossover region (points enclosed by circles in Fig.~\ref{fig:fund-phase-diagram}).
See Table~\ref{tab:zero-t-dataset} for the required zero-temperature data.
We plot $T_c$ versus the fermion mass $m_4$ in Fig.~\ref{fig:F-Tc-vs-mq}.%
\footnote{Note that our rough estimate of $T_c(m_4)$ does not correspond to the peak of any susceptibility in the crossover region.}

Here also, we expect that the first-order deconfinement transition of the pure gauge theory will reappear as the fermion mass rises towards infinity. Our simulations have not yet encountered this transition for masses as large as $m_{4} \approx 0.32 / \sqrt{t_1} = 250~\mathrm{MeV}$.
 
\begin{figure}[htb!]
	\centering
	\includegraphics{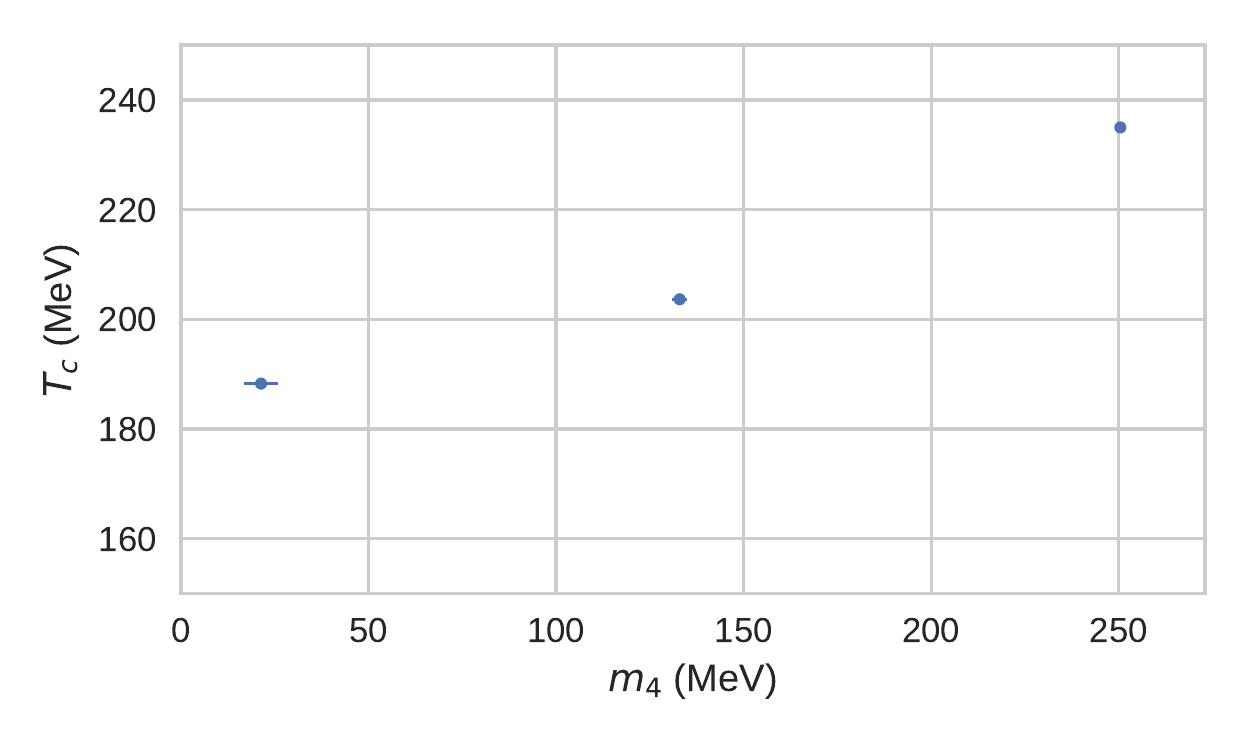}
	\caption{
		Temperature for the crossover of the fundamental-only theory with $N_t=6$ as a function of the AWI mass.
		The scale has been set via $t_1 \equiv 1/(780~\text{MeV})^2$.
	}
	\label{fig:F-Tc-vs-mq}
\end{figure}

\section{Phase Structure of the full theory}
   	\label{sec:phase}
   	\begin{figure}[htb]
	\includegraphics{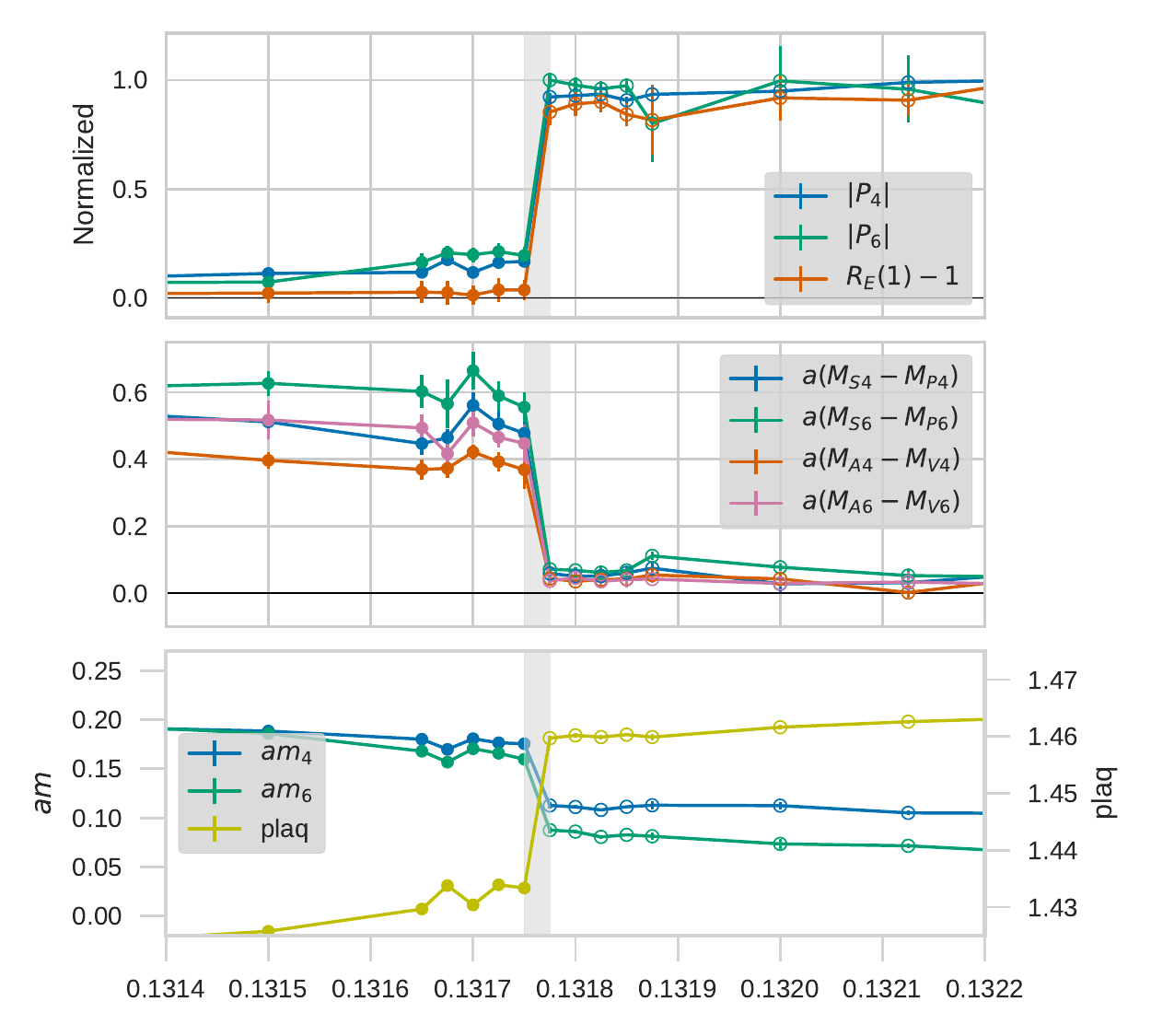}
	\caption{
		Behavior of various quantities in the full theory, varying $\kappa_6$ across the transition while holding $\beta=7.4$ and $\kappa_4=0.1285$ constant on $12^3 \times 6$.
		The gray band brackets the transition.
		Points with closed (open) circles are confined (deconfined) according to the behavior of Polyakov loops at long flow time.
		\textbf{Top:}
		Unflowed Polyakov loops for both representations and the flowed anisotropy.
		All diagnostics of confinement show simultaneous discontinuities.
		\textbf{Middle:}
		Mass splittings of parity partner mesonic states: scalar vs.~pseudoscalar, and vector vs.~axial vector.
		Chiral symmetry restoration occurs simultaneously for the two representations.
		\textbf{Bottom:}	
		AWI fermion masses for both representations, and the plaquette.
		All quantities jump discontinuously at the transition.
		}
	\label{fig:mrep-triple-slice}
\end{figure}

Figure~\ref{fig:mrep-triple-slice} depicts a slice through bare parameter space in the full theory with both fermion species dynamical, varying $\kappa_6$ while holding $\beta=7.4$ and $\kappa_4=0.1285$ fixed.
The top panel shows the behavior of our confinement diagnostics.
The Polyakov loops for both representations and the flowed anisotropy all jump simultaneously.
This means that the two species confine simultaneously.
The middle panel shows the behavior of the chiral diagnostics for both 
representations, the mass splittings of parity partner states.
The parity partners of both representations are split significantly at small $\kappa_6$, but simultaneously become nearly degenerate as $\kappa_6$ is increased.
This indicates that chiral symmetry restoration occurs simultaneously for the two representations.
Comparing the top and middle panels, we see that the combined confinement transition and the combined chiral transition coincide.
Away from the single jump, all quantities vary smoothly.
This behavior is typical throughout the region of bare parameter space that we have investigated: all phase diagnostics jump simultaneously and only once.
Thus, we find only two phases: a low-temperature phase where all fermions are confined and chirally broken and a high-temperature phase where all fermions are deconfined and chirally symmetric.

\begin{figure}[htb]
	\includegraphics[width=\linewidth,trim={0 1.75cm 0 2.25cm},clip]{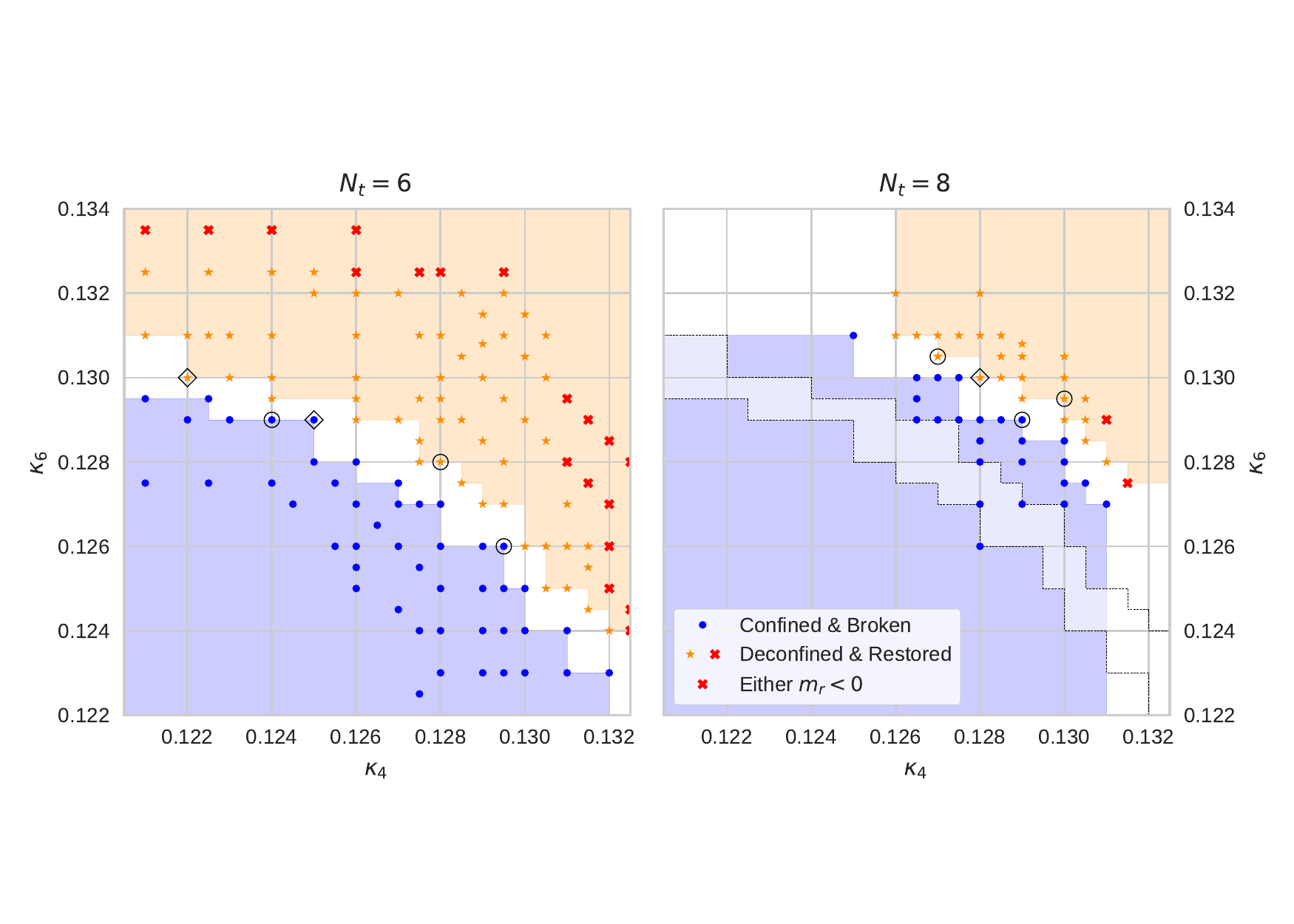}
	\caption{
		Phase diagram for $\beta=7.75$.
		To the left is the phase diagram for $N_t=6$ lattices, while to the right is the same region of bare parameter space for $N_t=8$ lattices.
		Blue dots indicate confined and chirally broken ensembles.
		Yellow stars indicate deconfined and chirally restored ensembles with $m_r>0$ for both species;
		red Xs are in regions where $m_4<0$ or $m_6<0$ or both.
		In the right figure, the transition region from $N_t=6$ is overlaid in gray, demonstrating that the transition moves as $N_t$ is varied.
		The circled ensembles have matching zero-temperature ensembles available.
		Ensembles enclosed by diamonds are where the phase changed when volume was changed (see text).
	}
	\label{fig:mrep-map-775}
\end{figure}

\begin{figure}[htb]
	\includegraphics[width=\linewidth,trim={0 0.5cm 0 1cm},clip]{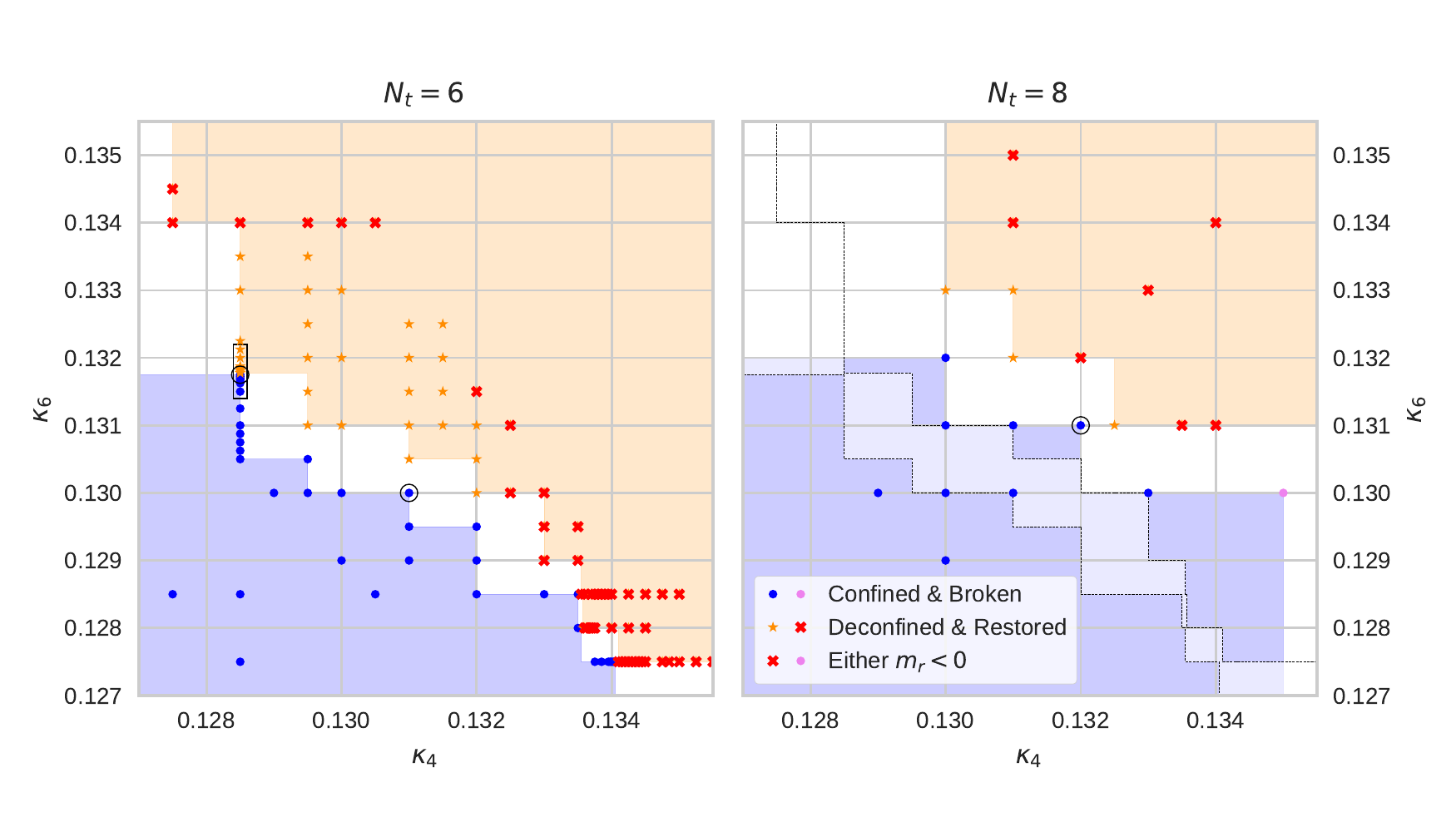}
	\caption{
		Phase diagram for $\beta=7.4$, similar to Fig.~\ref{fig:mrep-map-775}.
		The violet dot is a confined and chirally broken ensemble with $m_{4}<0$.
		The black box indicates the slice through bare parameter space shown in Fig.~\ref{fig:mrep-triple-slice}.
	}
	\label{fig:mrep-map-74}
\end{figure}

Figures~\ref{fig:mrep-map-775} and \ref{fig:mrep-map-74} show phase diagrams for the theory at $\beta=7.75$ and $\beta=7.4$ for $N_t=6$ and $N_t=8$.
In these plots, confined and chirally broken regions of parameter space are highlighted in blue while deconfined and chirally restored regions of the parameter space are highlighted in orange.
The transition thus lies somewhere in the white band in each phase diagram.
Points enclosed by diamonds are where the phase diagnostics change when varying $N_s$ from 12 to~18 when $N_t=6$ and from 16 to~24 when $N_t=8$.
The absence of many such points indicates that the location of the transition is insensitive to finite-volume effects.
Points enclosed by circles have zero-temperature data available at parameters near the transition (see Table~\ref{tab:zero-t-dataset}).
Comparing the left and right panels in Figs.~\ref{fig:mrep-map-775} and \ref{fig:mrep-map-74}, we see that the transition moves substantially in bare parameter space as we vary $N_t$.
This behavior is consistent with a thermal transition, and inconsistent with a bulk transition.

As shown in the bottom panel of Fig.~\ref{fig:mrep-triple-slice}, the plaquette and quark masses show a discontinuous jump at the transition, providing strong evidence that the observed transition is first order.
In support of this finding, we have also observed several tunneling events in the process of equilibrating new ensembles near the transition.
We have observed this to occur after more than $1000$ trajectories, much longer than the typical equilibration time for this volume.

We are interested in determining whether the transition temperature $T_c$ is comparable to its value in QCD, how strongly $T_c$ depends on fermionic effects, and how significant are lattice spacing artifacts in $T_c$.
In the full theory, $T_c$ is a function of the fermion masses and $1/N_t$.
We do not have sufficient data to constrain the location of the transition with any of $m_4$, $m_6$, or $a$ held fixed.
Instead, we examine the behavior of $T_c$ as we interpolate along the transition at fixed $\beta$ and $1/N_t$. 
For simplicity, we use $\kappa_4$ to parameterize each transition curve, along which all of $m_4$, $m_6$, and~$T_c$ vary.
We have estimated the lattice spacing and thus $T_c$ using the fit to $t_1/a^2$ described in Appendix~\ref{sec:scale-setting}.
Tracing along the transition bands in Figs.~\ref{fig:mrep-map-775} and~\ref{fig:mrep-map-74}, wherever there are ensembles matched in $\kappa_4$ on the edges of the transition band, we use our $t_1/a^2$ fit to estimate upper and lower bounds for $T_c$.
The resulting bounds on $T_c$ as a function of $\kappa_4$ are the horizontal black dashes with error bands in Figs.~\ref{fig:Tc-beta-74} and~\ref{fig:Tc-beta-775}.
As indicated by the spanning arrows, the transition temperature must lie between these bounds.
Note that there is an uncontrolled systematic error for these bounds: the phases of some ensembles near the transition edge may be misdiagnosed if they have not been equilibrated long enough to tunnel to the correct phase.

As in the sextet-only and fundamental-only theories, $T_c$ is comparable to its value in QCD (150~MeV) and SU(3) pure gauge theory (280~MeV).
For both $N_t=6$ curves, $T_c$ may not remain constant as we vary $\kappa_4$ to interpolate along the transition.
We cannot exclude that the dependence may be a lattice artifact, but as one might expect from fermionic influence on the transition, $T_c$ appears to depend more strongly on $\kappa_4$ at $\beta=7.75$ than at $\beta=7.4$.
Comparing with the $\kappa_c$ curves of Fig.~19 of Ref.~\cite{Ayyar:2017qdf}, we see that at $\beta=7.4$ the transition curve is roughly parallel to $\kappa_c$ and thus traces lines of approximately constant quark mass for whichever species is lighter; this slow variation in the masses is consistent with the observed slow variation in $T_c$. Meanwhile, at $\beta=7.75$ the transition curve moves further away from $\kappa_c$ when $\kappa_4 \approx \kappa_6$, leading to heavier fermions; when the fermions are heavier, the system becomes more pure-gauge-like and $T_c$ increases.

\begin{figure}[htb]
	\includegraphics[width=\linewidth]{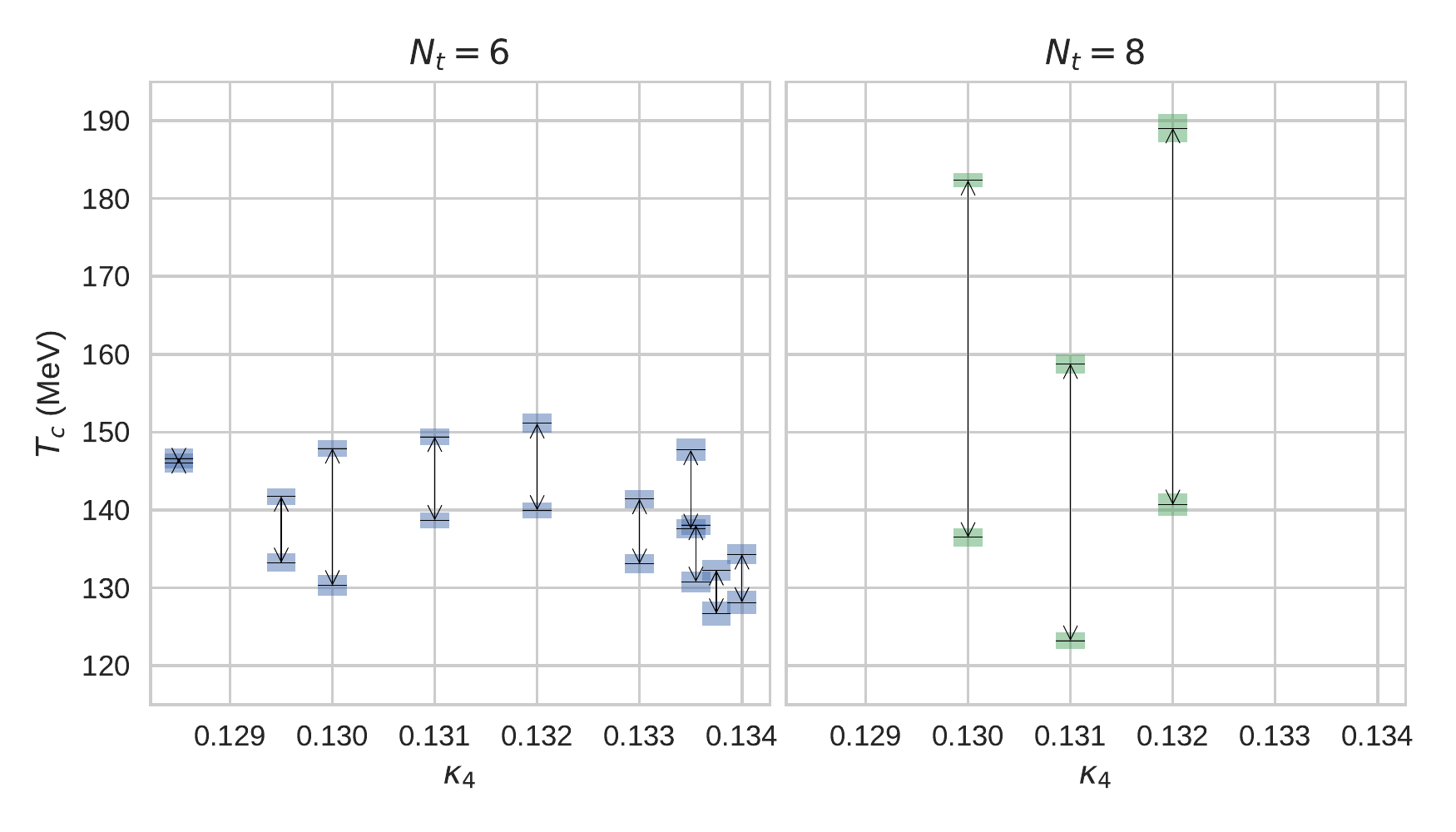}
	\caption{
		Transition temperature of the multirep theory at $\beta=7.4$ as a function of $\kappa_4$ on $N_t=6$ and $N_t=8$.
		Lattice spacings used to determine the temperature are computed using the fit to $t_1/a^2$ with $t_1 = 1/(780~\text{MeV})^2$ as discussed in Appendix~\ref{sec:scale-setting}.
		Axes are matched between $N_t=6$ and $N_t=8$.
		Black lines with error bands indicate the temperature on ensembles on either end of the transition bands in Fig.~\ref{fig:mrep-map-74}.
		The transition temperature lies in the span indicated by the double-headed arrows.
	}
	\label{fig:Tc-beta-74}
\end{figure}

\begin{figure}[htb]
	\includegraphics[width=\linewidth]{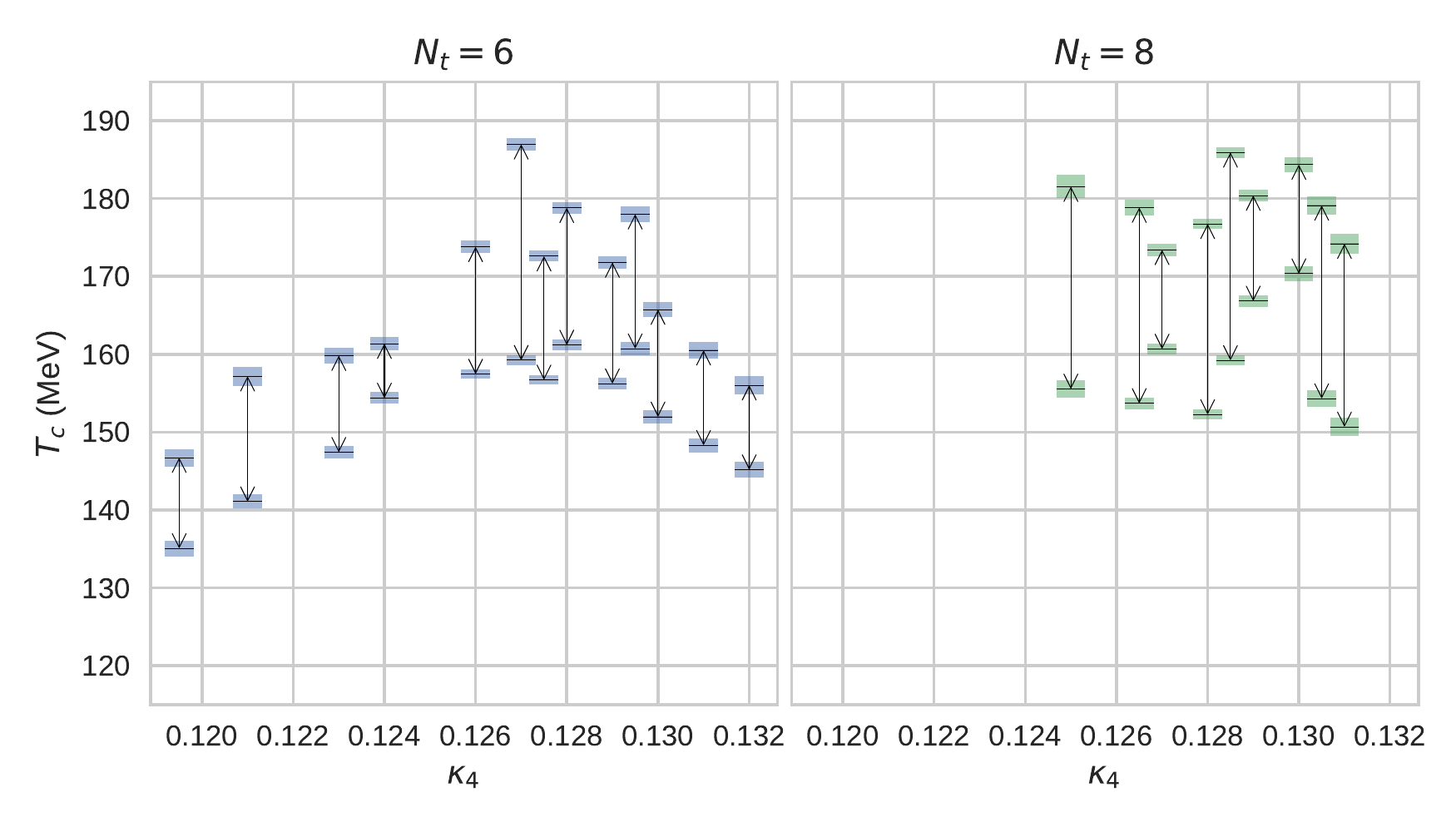}
	\caption{
		Transition temperature of the multirep theory at $\beta=7.75$ as a function of $\kappa_4$ on $N_t=6$ and $N_t=8$.
		Similar to Fig.~\ref{fig:Tc-beta-74}.
	}
	\label{fig:Tc-beta-775}
\end{figure}

\section{Conclusions}
	\label{sec:conclusions}
    	We summarize our results for the phase structure of our theory in two Columbia plots  in Fig.~\ref{fig:col_plot}.
The axes in Fig.~\ref{fig:col_plot_qmass_lat} are plotted in lattice units, while the axes in Fig.~\ref{fig:col_plot_qmass_mev} are plotted in physical units, as defined in Appendix~\ref{sec:scale-setting}.
Each plot has two kinds of symbols.
Open symbols show quark masses from matching zero-temperature simulations: $a m_{4}$, $a m_{6}$, and $t_1/a^2$ are all taken from zero-temperature ensembles run at bare parameters matched to finite-temperature ensembles near the transition.
Closed symbols show the quark masses from finite-temperature simulations: $am_{4}$ and $a m_{6}$ are measured on finite-temperature ensembles along the deconfined side of the transition curve, and $t_1/a^2$ is obtained from the fit described in Appendix~\ref{sec:scale-setting}.
If there were no lattice artifacts, the open and closed symbols for each $(N_t, \beta)$ set would coincide.
The solid lines in Fig.~\ref{fig:col_plot_qmass_mev}, which come from the interpolation, lie close to the open symbols, which were measured directly and did not require a fitting function.
The curves in Fig.~\ref{fig:col_plot} where $m_{4} \ne \infty$ and $m_{6} \ne \infty$ indicate the masses at which we have examined the phase structure of the full theory in detail and found only a single first-order thermal transition.
The points where $m_{4}=\infty$ indicate where we have examined the sextet-only theory and found only a single transition.
The points where $m_{6}=\infty$ indicate where we have examined the fundamental-only theory and found only a single crossover or transition.

\begin{figure}[htb!]
\centering    
\subfigure[~Columbia plot with quark masses in lattice units]{
	\label{fig:col_plot_qmass_lat}
	\includegraphics[width=10cm,clip,trim=0.1in 0.5in 0.6in 0.75in]{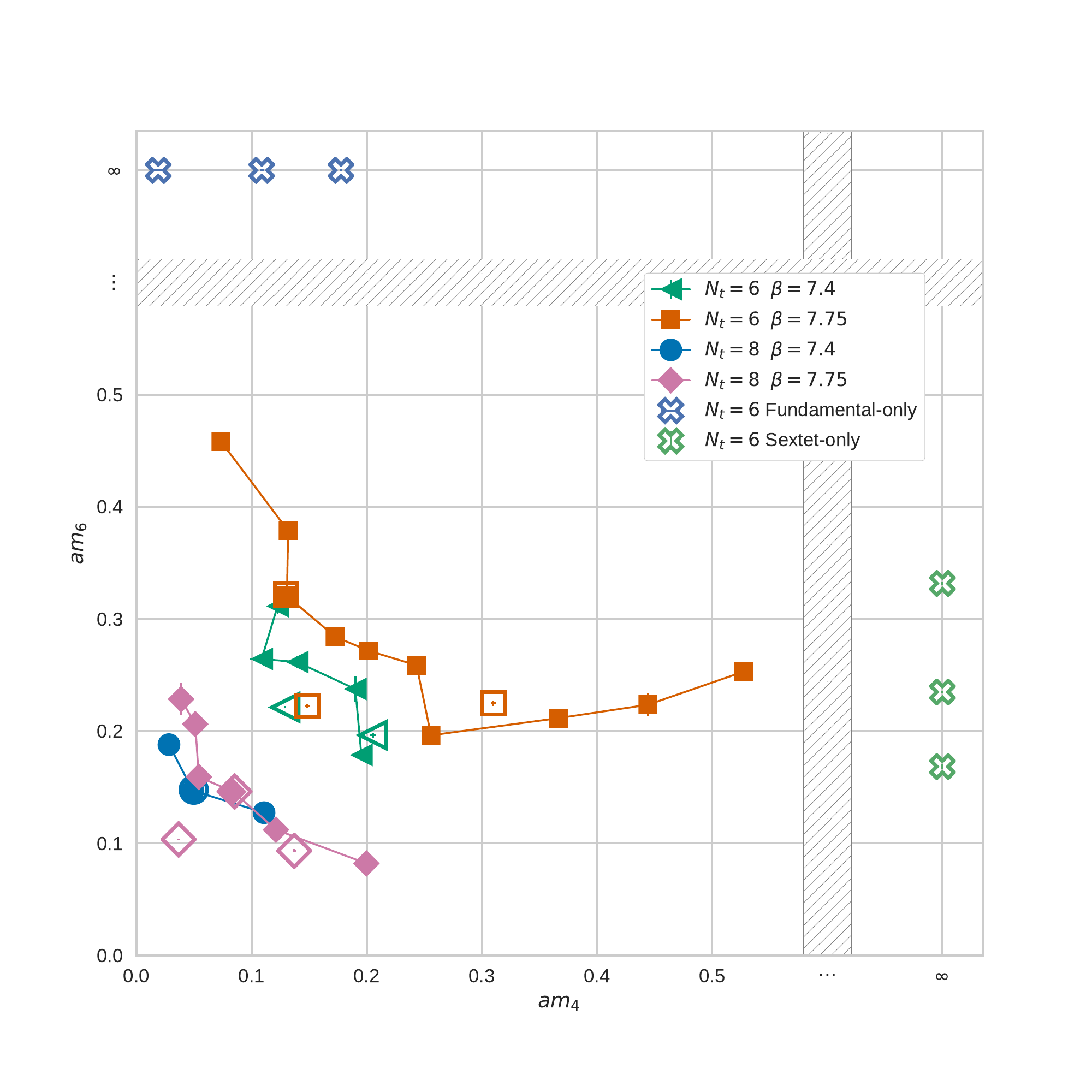}
}
\subfigure[~Columbia plot with quark masses in MeV]{
	\label{fig:col_plot_qmass_mev}
	\includegraphics[width=10cm,clip,trim=0.1in 0.5in 0.6in 0.75in]{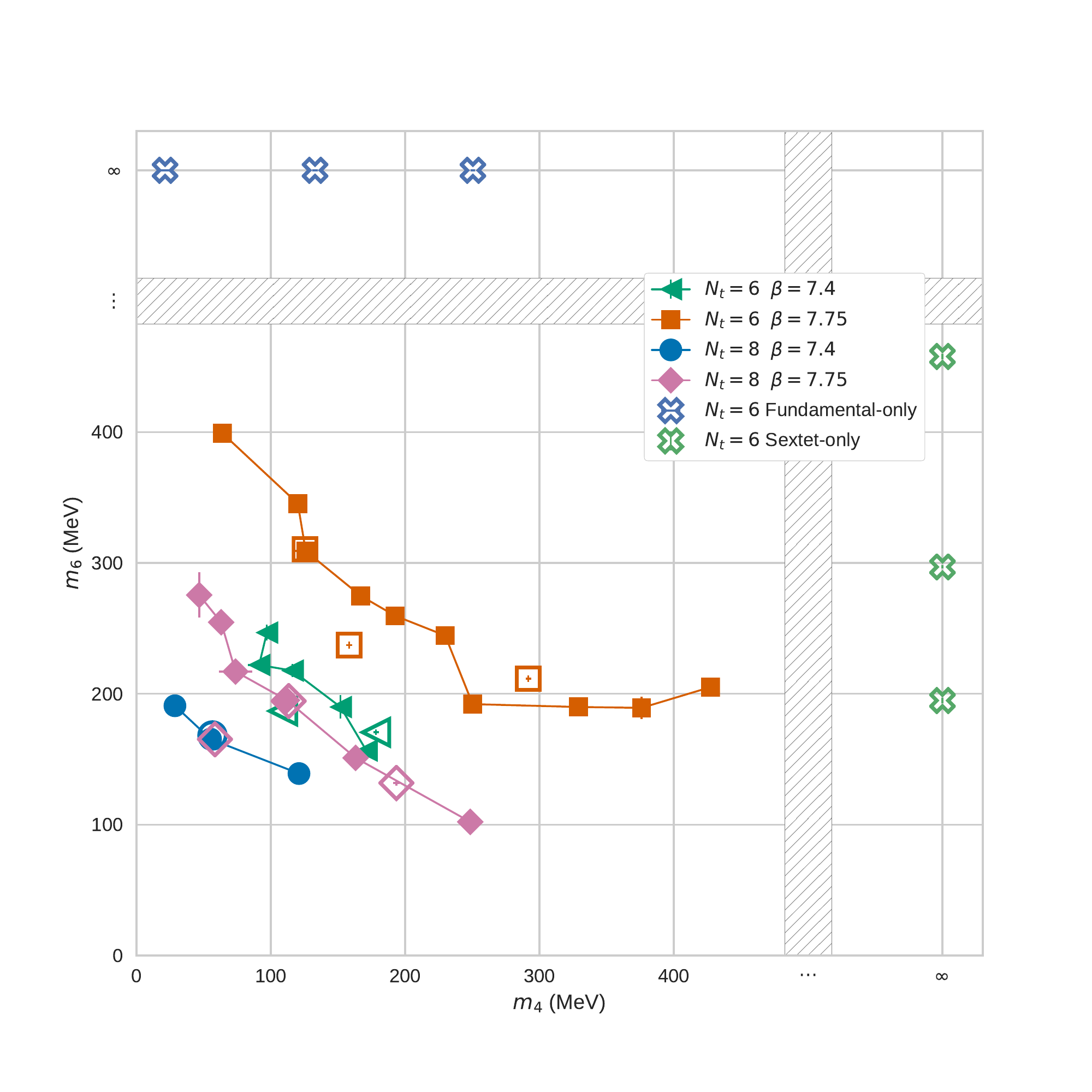}
}
\caption{
	Columbia plots, by analogy with QCD\@.
	In Fig.~\ref{fig:col_plot_qmass_lat}, the quark masses are in lattice units.
	In Fig.~\ref{fig:col_plot_qmass_mev}, quark masses are in MeV [$t_1 \equiv 1/(780~\text{MeV})^2 $], as defined in 
	Appendix~\ref{sec:scale-setting}.
	Each color and symbol is associated with a different $\beta$ and $N_t$.
	Closed symbols are finite-temperature quark masses, while hollow symbols are zero-temperature quark masses from ensembles on the transition boundary
	The lattice spacings for the zero temperature quark masses are computed directly from $t_1/a^2$ on that ensemble. The lattice spacings for the finite-temperature quark masses are computed from the model described in Appendix~\ref{sec:scale-setting}.
}
\label{fig:col_plot}
\end{figure}

Our numerical investigation of the full multirep theory finds only a single, first-order thermal transition.
This non-observation of separated chiral phase transitions is in direct contradiction to predictions of the Most Attractive Channel hypothesis, according to which the sextet fermions should condense before the fundamentals as the temperature is lowered.
While our exploration of the three-dimensional parameter space is by no means exhaustive, and separated phase transitions might exist for some values of fermion mass, we have examined the theory at masses ranging from 50 to 400~MeV and ruled them out in this domain.
We similarly find that the fundamental-only and sextet-only theories appear to be QCD-like, with a combined chiral and confinement transition.

In the multirep theory we find a strongly first-order phase transition.
This is consistent with a one-loop Pisarski-Wilczek scaling analysis
appropriate to the limit where ${m_4 = m_6 = 0}$ \cite{Hackett:2017gti}.

Increasing the quark masses in the single-species theories, we expect eventually to run into the first-order region of the pure-gauge transition.
Our exploration of these theories stops short of the quark masses at which this occurs.
Similarly, because we have only explored relatively light quark masses, we are unable to determine whether the first-order regions surrounding the double-chiral limit and the pure-gauge limit are connected.

We observe no bulk transitions in the full  theory or in either of its single-species limits.
We attribute this to our use of the NDS action.
Comparison to the previous study of the sextet-only theory \cite{DeGrand:2015lna} makes this clear.

The finite-temperature properties of this model may have important implications for cosmology
if it or something like it happens to be realized in nature.
First order transitions in the early Universe give rise to gravitational 
waves with distinctive properties \cite{Schwaller:2015tja,Caprini:2015zlo}.
These signals may be accessible to near-future gravitational wave detectors such as LISA \cite{Caprini:2015zlo}.

\begin{acknowledgments}

Research was supported by U.S.~Department of Energy Grant
Number  under grant DE-SC0010005 (Colorado)
and  by the Israel Science Foundation under grants 449/13 and 491/17.
Brookhaven National Laboratory is supported
by the U.~S.~Department of Energy under contract DE-SC0012704.
Some computations were performed on the University of Colorado cluster.
This work utilized the Janus supercomputer, which is supported by the National Science Foundation (award number CNS-0821794) and the University of Colorado Boulder. The Janus supercomputer is a joint effort of the University of Colorado Boulder, the University of Colorado Denver and the National Center for Atmospheric Research.
Additional computations were done on facilities of the USQCD Collaboration at Fermilab,
which are funded by the Office of Science of the U.~S. Department of Energy.
The computer code is based on the publicly available package of the
 MILC collaboration~\cite{MILC}.

\end{acknowledgments}

\appendix

\section{Scale setting and transition temperature}
	\label{sec:scale-setting}
	
\subsection{Setting the scale using Wilson flow}

In order to estimate the transition temperature, we need the lattice spacing.
To this end, we generated zero-temperature ensembles at points in bare parameter space along the transition surface and measured the Wilson flow scale $t_1$ \cite{Sommer:2014mea}.

We can define a general flow scale $t_*$ via
\begin{equation}
\ev{t^2 E(t)}|_{t=t_*} = C
\label{eqn:Cdefn}
\end{equation}
where $E(t)$ is a discretization of the gluonic action density $\Tr G_{\mu \nu} G^{\mu \nu}$ at flow time $t$, and $C$ is a constant defining the scale.
In QCD, for $t_0$ one sets $C = 0.3$ and for $t_1$ one sets $C = 2/3$; the physical scale enters by fixing
$\sqrt{t_0} \simeq 0.142$~fm~\cite{Bazavov:2015yea}.
To translate quantities to physical values for comparison with QCD, we need some way of matching our definition of $t_0$ in $N_c=4$ with the definition in $N_c=3$.

Large-$N_c$ scaling arguments provide a translation procedure \cite{DeGrand:2017gbi,hudspith-2017-bsmtalk}.
The observable $\ev{t^2 E(t)}$ can be used to define the renormalized coupling at the scale $t$ \cite{Ce:2016awn,Vera:2017dxr},
\begin{equation}
g^2_\text{wf}(t)\equiv\frac{128 \pi^2}{3 (N_c^2-1)} \ev{t^2 E(t)}  .
\label{eqn:wf-coupling-def}
\end{equation}
The usual large-$N_c$ scaling argument holds the 't~Hooft coupling $\lambda_0 = g_0^2 N_c$ constant in $N_c$.
Thus, $g^2 \sim 1/N_c$ at leading order, regardless of renormalization scheme.
We can immediately read off from Eq.~(\ref{eqn:wf-coupling-def}) that, for $g^2 \sim 1/N_c$ to hold, $\ev{t^2 E(t)} \sim N_c$ at leading order.
Comparing Eq.~(\ref{eqn:Cdefn}), we conclude that a reasonable choice is $C\propto N_c,$
and so for SU(4) we define the scales $t_0$ and $t_1$ via
\begin{align}
	\ev{t_0^2 E(t_0)} &= 0.4 \\
	\ev{t_1^2 E(t_1)} &= \frac{8}{9}.
\end{align}

Assuming that our $N_c$ scaling has produced an equivalent quantity in SU(4), we may take the value $t_0 = (0.142~\mathrm{fm})^2$ from SU(3) \cite{Bazavov:2015yea}.
Our data set, however, samples the transition at relatively large lattice spacings $a \gtrsim \sqrt{t_0}$.
In this regime, $t_0$ develops nonlinear lattice-spacing artifacts; in practice, this is the failure of $\ev{t^2 E(t)}$ to reach the linear regime before it exceeds 0.4, and so we are sampling the ``knee'' in the typical $\ev{t^2 E}$ trajectory.
Because $t_1$ is larger, it can be measured on larger lattice spacings than $t_0$.
In order to use $t_1$ in lieu of $t_0$, we need a conversion factor.

In Fig.~\ref{fig:t1-vs-t0} we examine the behavior of the ratio of $\sqrt{t_1/t_0}$ as a function of the lattice spacing over our entire zero-temperature data set \cite{Ayyar:2017qdf}.
The quantities $\sqrt{t_1}$ and $\sqrt{t_0}$ are fixed lengths, so their ratio should be some constant independent of lattice spacing.
We see, however, that at large lattice spacing the ratio is not constant.
Making the cut $a < 1/\sqrt{2 t_1}$ (the dashed line in Fig.~\ref{fig:t1-vs-t0}), we find $\sqrt{t_1/t_0} \simeq 1.77$, and so $\sqrt{t_1} \simeq 0.252~\mathrm{fm}$ or equivalently $1/\sqrt{t_1} \simeq 783~\mathrm{MeV}$.
For simplicity we take ${1/\sqrt{t_1} \equiv 780~\mathrm{MeV}}$ for our fiducial value.
Determining the physical temperature of any given ensemble merely amounts to measuring the dimensionless quantity $t_1/a^2$ in order to calculate $T=(N_ta)^{-1}=(N_t\sqrt{t_1})^{-1}\times\sqrt{t_1/a^2}$.

\begin{figure}
	\includegraphics[width=3.5in]{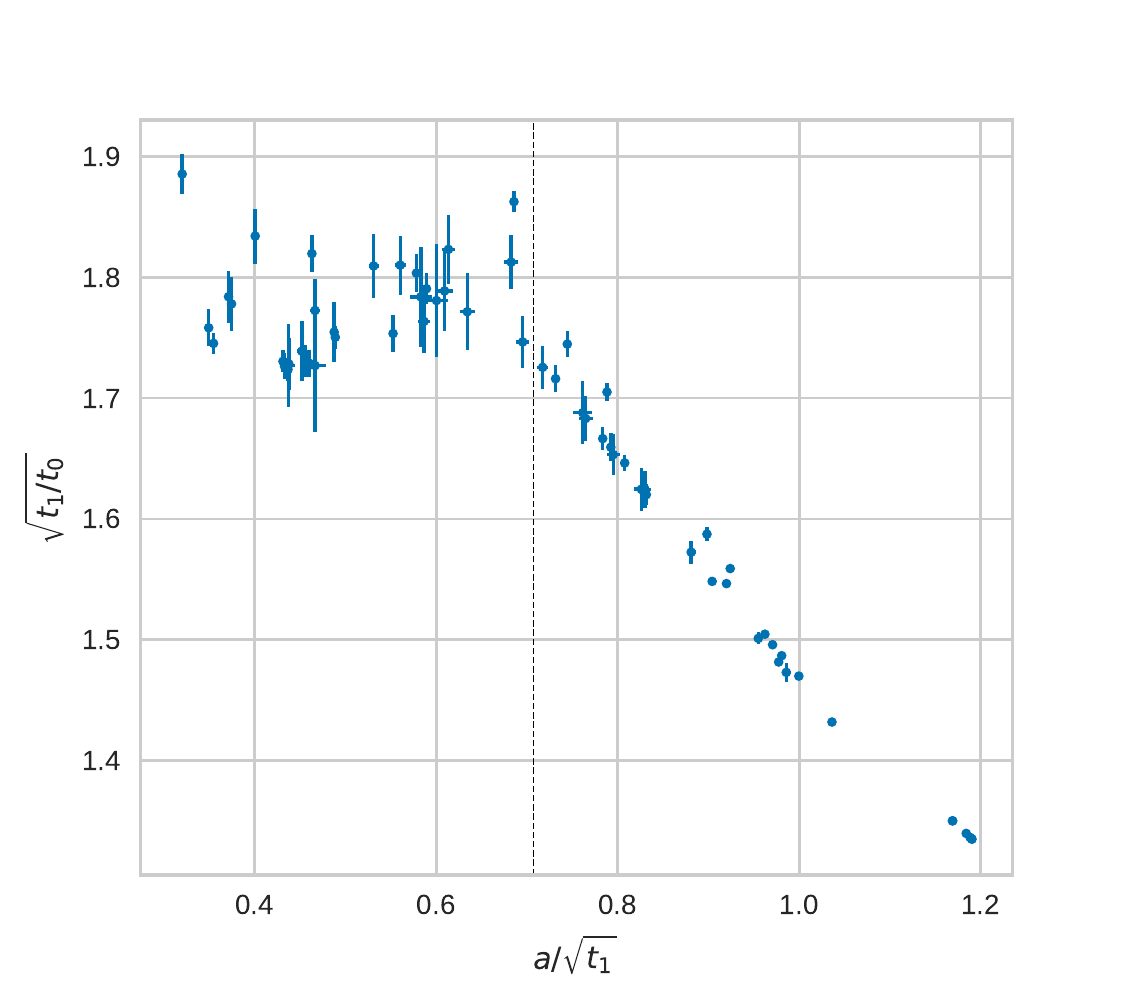}
	\caption{
		Ratio of the Wilson flow scales $\sqrt{t_1}$ and $\sqrt{t_0}$ as a function of lattice spacing, plotted against lattice spacing measured in units of $\sqrt{t_1}$.
		Above $a \sim 0.8$, large lattice spacing effects begin to contaminate $t_0$.
	}
	\label{fig:t1-vs-t0}
\end{figure}

\subsection{Fitting the lattice spacing}

We have only a few zero-temperature ensembles in the region of bare parameters relevant to the transition.
In order to perform a more detailed temperature analysis, we modeled $t_1/a^2$ as a function of the bare parameters.
We do not have any theoretical expectations for the form of $t_1/a^2$, so this modeling is entirely empirical.

To include in our fit, we have 39 zero-temperature ensembles where $t_1/a^2 > 1$ to avoid discretization effects; of these, 12 ensembles are at $\beta=7.75$.
Lattice spacings computed using the Wilson flow are easily determined to very high statistical precision.
Without knowing the true form of $t_1/a^2$ as a function of $\beta$, $\kappa_4$, and~$\kappa_6$, and with our limited dataset, we are unable to produce a model that can fit $t_1/a^2$ with convincingly small $\chi^2$.
We are thus led to inflate the errors on $t_1/a^2$ to 2.5\% of the value.
Such a model predicts the value of $t_1/a^2$ to within $2.5\%$ for any $(\beta,\kappa_4,\kappa_6)$ in the region of interest.

The form of our model is motivated by several observations about the behavior of $t_1/a^2$ as a function of the bare parameters $(\beta, \kappa_4, \kappa_6)$.
At fixed $\beta$, we observe that (1) as $\kappa_4$ or $\kappa_6$ is increased, $t_1/a^2$ increases monotonically; (2) $t_1/a^2$ varies smoothly as a function of $\kappa_4/\kappa_6$, in such a way that there are smooth curves of constant $t_1/a^2$; (3) these curves of constant $t_1/a^2$ are roughly elliptical in shape; and (4) as both $\kappa$'s go to zero and the fermions decouple, $t_1/a^2$ settles to a constant.
This motivates the functional form
\begin{equation}
	\frac{t_1}{a^2} = \exp \left[ \frac{r^2 - r_0^2(\beta)}{\gamma(\beta)} \right]
	+ C(\beta)\,,
	\label{eqn:t1-model}
\end{equation}
where
\begin{equation}
	r^2 \equiv (8\kappa_4)^2 + \alpha(\beta) (8\kappa_6)^2\\
\end{equation}
and $\alpha(\beta)$, $\gamma(\beta)$, $r_0^2(\beta)$, and $C(\beta)$ are functions of $\beta$ only and thus constant at fixed $\beta$.
An equally reasonable functional form would be a power law rather than an exponential; however, fits to such functional forms produce unreasonably large powers and do not model the data as well.
Note that we write the model in terms of $8\kappa_4$, $8\kappa_6$, and $\beta/8$ (below) which are all $O(1)$, so that the size of the fit parameters may be compared easily.
Physically, $r_0^2$ quantifies the value of $r^2$ where fermionic effects are frozen out; $\alpha$ projects the elliptical curves of constant $t_1/a^2$ in the $\kappa_4$--$\kappa_6$ plane to circles; $\gamma$ quantifies how quickly $t_1/a^2$ increases as fermionic effects become strong; and $C$ is the value of $t_1/a^2$ in the pure-gauge limit where $\kappa_4=\kappa_6=0$ [up to a small $\exp(-r_0^2/\gamma)$ correction].

To obtain a concrete realization of the abstract model (\ref{eqn:t1-model}), we approximate $\alpha(\beta)$, $\gamma(\beta)$, and $r_0^2(\beta)$ as linear functions,
\begin{equation}
	\begin{aligned}
		\alpha(\beta) &= \alpha_0 + \alpha_1 \left(\frac{\beta}{8}\right) \\
		\gamma(\beta) &= \gamma_0 + \gamma_1 \left(\frac{\beta}{8}\right) \\
		r_0^2 &= R_0 + R_1 \left(\frac{\beta}{8}\right)
	\end{aligned}
	\label{eqn:t1-model-linear-fns}
\end{equation} 
where $\alpha_0$, $\alpha_1$, $\gamma_0$, $\gamma_1$, $R_0$, and $R_1$ are fit parameters.
In the pure-gauge limit where $\kappa_4=\kappa_6=0$, as $\beta \rightarrow 0$, we expect $a \rightarrow \infty$ and thus $t_1/a^2 \rightarrow 0$.  We thus demand that $C(0)=0$, and model $C(\beta)$ as a power,
\begin{equation}
	C(\beta) = C_0 \left(\frac{\beta}{8}\right)^{C_1},
	\label{eqn:t1-model-power-law}
\end{equation}
where $C_0$ and $C_1$ are fit parameters.
Fitting the 39 ensembles of our zero-temperature $t_1/a^2$ dataset to the model defined by Eqs.~(\ref{eqn:t1-model})--(\ref{eqn:t1-model-power-law}), we obtain the fit parameters in Table~\ref{tab:t1-fit-params} with $\chi^2/31=0.87$ and $Q=1-P=0.67$.
The resulting model predicts the value of $t_1/a^2$  within $5\%$ for all 39 ensembles included in the fit, and within $2.5\%$ for 28 of these ensembles.
Figure~\ref{fig:t1-fit-preds} shows the predictions of the model versus data at $\beta=7.75$.

\begin{table}
	\centering
	\begin{ruledtabular}\begin{tabular}{cdcd}
		Parameter &  & Parameter &  \\\hline
		$\alpha_0$ & 6.1(8) & $\alpha_1$ & -4.6(8) \\
		$\gamma_0$ & -0.48(6) & $\gamma_1$ & 0.56(6) \\
		$R_0$ & 10.7(9) & $R_1$ & -8.2(9) \\
		$C_0$ & 0.94(7) & $C_1$ & 2.4(1.6)\\
	\end{tabular}\end{ruledtabular}
	\caption{
		Best-fit parameters for the $t_1/a^2$ model defined by Eqns.~\ref{eqn:t1-model}, \ref{eqn:t1-model-linear-fns}, and \ref{eqn:t1-model-power-law}.
		}
	\label{tab:t1-fit-params}
\end{table}

\begin{figure}[htb]
	\includegraphics{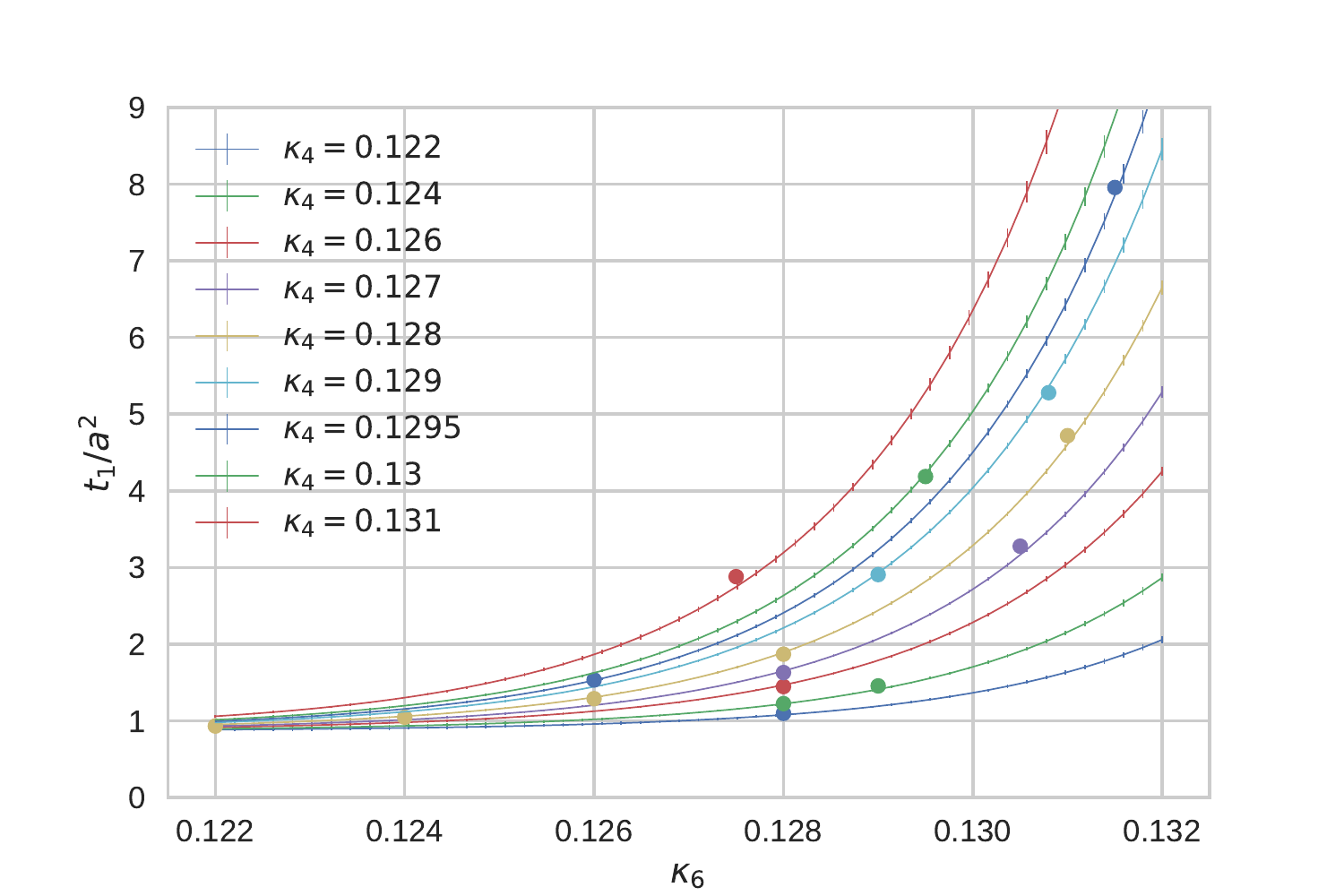}
	\caption{
		Lines are predictions for $t_1/a^2$ as a function of $\kappa_6$ for various $\kappa_4$ at $\beta=7.75$ by the model defined by Eqs.~(\ref{eqn:t1-model})--(\ref{eqn:t1-model-power-law}) and the best-fit parameters of Table~\ref{tab:t1-fit-params}.
		Dots are $t_1/a^2$ data at $\beta=7.75$.
		Colors are matched between dots and lines at the same $\kappa_4$.
	}
	\label{fig:t1-fit-preds}
\end{figure}

\begin{table}
	\centering
	\begin{ruledtabular}
		\begin{tabular}{rll}
			& Fit to $\beta=7.75$ only &  Full fit at $\beta=7.75$  \\\hline
			$\alpha(7.75)$ & 1.61(5)   &  1.58(4) \\
			$\gamma(7.75)$ & 0.062(2)  &  0.061(2) \\
			$r_0^2(7.75)$  & 2.73(5)   &  2.70(4) \\
			$C(7.75)$      & 0.87(3)   &  0.87(3)\\
		\end{tabular}
		\end{ruledtabular}
		\caption{
			Model parameters at $\beta=7.75$ from direct fit to $\beta=7.75$ data versus predictions for those parameters from Eqs.~(\ref{eqn:t1-model-linear-fns}) and~(\ref{eqn:t1-model-power-law}), and the best-fit values in Table~\ref{tab:t1-fit-params}.
		}
		\label{tab:t1-fit-comparison}
	\end{table}

To cross-check our model, we compare with fits of subsets of the dataset to simpler models.
At fixed $\beta$, all of $\alpha(\beta)$, $\gamma(\beta)$, $r_0^2(\beta)$, and $C(\beta)$ are constant, providing a simplified four-parameter realization of Eq.~(\ref{eqn:t1-model}).
Fitting to 12 ensembles at fixed $\beta=7.75$ yields the model parameters in the second column of Table~\ref{tab:t1-fit-comparison} with $\chi^2/12=0.78$ and $Q=1-P=0.67$.
By comparison, the model parameters in the third column of Table~\ref{tab:t1-fit-comparison} are predictions obtained by plugging the best-fit parameters in Table~\ref{tab:t1-fit-params} into Eqs.~(\ref{eqn:t1-model-linear-fns}) and~(\ref{eqn:t1-model-power-law}) for $\beta=7.75$.
Even though the full-dataset fit includes more than three times as many ensembles, the parameters agree closely; if this did not hold, it would suggest that the model is overfitting the data.
For the 12 ensembles at $\beta=7.75$, the predictions of these two fits agree within $1.5\%$.

	\newpage

\clearpage

\section{Summary of ensembles near the transition}
	\label{sec:data}

\begin{table}[htb!]
	\centering
	\begin{ruledtabular}
		\begin{tabular}{llllllllll}
			$ N_s$ & $N_t$ & $\beta$ & $ \kappa_4 $  & $ \kappa_6 $  & $ { t_1 / a^2} $ & $ M_{P4} / M_{V4}$ & $ M_{P6} / M_{V6}$ & $\sqrt{t_1} m_{4}$ & $\sqrt{t_1} m_{6}$ \\ \hline
			16 &    18 &     9.4 &         0.123 &             - &        3.27(2) &            0.85(1) &                    - &          0.321(2) &                   - \\
			16 &    18 &     9.2 &         0.126 &             - &       2.460(8) &           0.742(8) &                    - &          0.171(3) &                   - \\
			16 &    18 &     9.0 &          0.13 &             - &        2.10(1) &            0.39(2) &                    - &          0.027(6) &                   - \\
			16 &    18 &     9.2 &             - &         0.115 &        3.40(3) &                    - &           0.940(9) &                   - &          0.999(5) \\
			16 &    18 &     9.0 &             - &        0.1205 &        3.13(2) &                    - &           0.896(4) &                   - &          0.587(3) \\
			16 &    18 &     8.8 &             - &         0.124 &        2.63(2) &                    - &           0.842(8) &                   - &          0.381(2) \\
			16 &    18 &     8.6 &             - &         0.127 &        2.20(2) &                    - &            0.79(1) &                   - &          0.250(3) \\
			16 &    32 &     7.4 &         0.131 &          0.13 &       1.171(3) &           0.765(5) &           0.824(3) &         0.140(1) &          0.240(1) \\
			16 &    18 &     7.4 &         0.132 &         0.131 &        2.13(2) &            0.60(3) &            0.78(1) &          0.073(2) &          0.216(3) \\
			12 &    24 &     7.4 &        0.1285 &       0.13175 &       1.239(8) &            0.83(1) &            0.79(1) &          0.229(3) &          0.219(3) \\
			16 &    18 &    7.75 &        0.1295 &         0.126 &        1.53(1) &            0.74(2) &           0.87(1) &          0.161(3) &          0.398(4) \\
			12 &    24 &    7.75 &         0.128 &         0.128 &        1.87(2) &            0.80(1) &           0.82(1) &          0.203(3) &          0.304(3) \\
			16 &    18 &    7.75 &         0.124 &         0.129 &       1.455(8) &           0.888(8) &            0.82(1) &          0.374(3) &          0.271(3) \\
			12 &    24 &    7.75 &         0.129 &         0.129 &        2.91(6) &            0.74(2) &            0.79(2) &          0.145(4) &          0.249(5) \\
			16 &    32 &    7.75 &          0.13 &        0.1295 &        4.19(4) &            0.61(2) &           0.787(7) &         0.0749(8) &          0.212(1) \\
			12 &    24 &    7.75 &         0.127 &        0.1305 &        3.28(4) &            0.82(2) &            0.73(2) &          0.248(3) &          0.169(3) \\
		\end{tabular}
	\end{ruledtabular}
	\caption{Zero-temperature data sets used to compute the scale near the thermal transition.}
	\label{tab:zero-t-dataset}
\end{table}

\bibliography{su4_finiteT}

\end{document}